\documentclass[10pt,journal,compsoc]{IEEEtran}

\pdfoutput=1

\ifCLASSOPTIONcompsoc
  
  \usepackage[nocompress]{cite}
\else
  \usepackage{cite}
\fi

\usepackage{amsmath}
\usepackage{csquotes}
\usepackage{multirow}
\usepackage{pbox}
\usepackage{braket}
\usepackage{graphics}
\usepackage{graphicx}
\usepackage{url}
\usepackage{comment}
\usepackage{booktabs}
\usepackage[ruled,linesnumbered]{algorithm2e}
\usepackage{balance}
\usepackage{xcolor}
\usepackage{multirow}
\ifCLASSINFOpdf
\else
\fi

\begin{document}

\title{Detect and Classify IoT  Camera Traffic}

\author{Priyanka Rushikesh Chaudhary 
        and Rajib Ranjan Maiti}

\IEEEtitleabstractindextext{%
\begin{abstract}
Deployment of IoT cameras in an organization threatens security and privacy policies, and the classification of network traffic without using IP addresses and port numbers has been challenging. 
In this paper, we have designed, implemented and deployed a system called \emph{iCamInspector} to classify network traffic arising from IoT camera in a mixed networking environment.
We have collected a total of about 36GB of network traffic containing video data from three different types of applications (four online audio/video conferencing applications, two video sharing applications and six IoT camera from different manufacturers) in our IoT laboratory. We show that with the help of a limited number of flow-based features, iCamInspector achieves an average accuracy of more than 98\% in a 10-fold cross-validation with a false rate of about 1.5\% in testing phase of the system. 
A real deployment of our system in an unseen environment achieves a commendable performance of detecting IoT camera with an average detection probability higher than 0.9.
\end{abstract}

}

\maketitle

\IEEEdisplaynontitleabstractindextext

\IEEEpeerreviewmaketitle

\IEEEraisesectionheading{\section{Introduction}\label{sec:introduction}}

A number of recent market studies have revealed that the smart cameras for indoor and/or outdoor surveillance market in 2020 has reached a huge volume of USD 3 billions which is about 70\% of all the IoT products in market \cite{market1Cam_2020,market2cam_2019}.  
Huge deployment of such cameras pose direct security and privacy threats to the owners and their associates both. 
Recent botnet attacks, like Parsirai \cite{Tim2017Blog} and Mirai \cite{manos2017Usenix}, have revealed the severity of security risks of such IoT devices.
Therefore, rather than focusing on patching security loopholes, the detection of IoT camera via either passive monitoring of network traffic or active fingerprinting has been an attractive research area in recent times \cite{wang2020iot},\cite{saidi2020imc},\cite{alharbi2018IET}. 
Supervised or unsupervised machine learning techniques turn out to be an effective weapon to achieve such objectives possibly to develop a better access control mechanism \cite{visoottiviseth2020IEEE},\cite{chaabouni2019IEEE},\cite{Snort},\cite{Suricata}. 
Use of machine learning is intensified because the IoT cameras tend to utilize the popular cloud services like AWS cloud\cite{IP_awscloud} making rule-based defense ineffective.  
In this paper, we aim to explore flow-based network traffic characteristics to classify network traffic containing video data of IoT (Spy) camera in an environment where traditional online audio/video applications are unavoidable.

\begin{figure}[h]
\centering
\includegraphics[width=\linewidth]{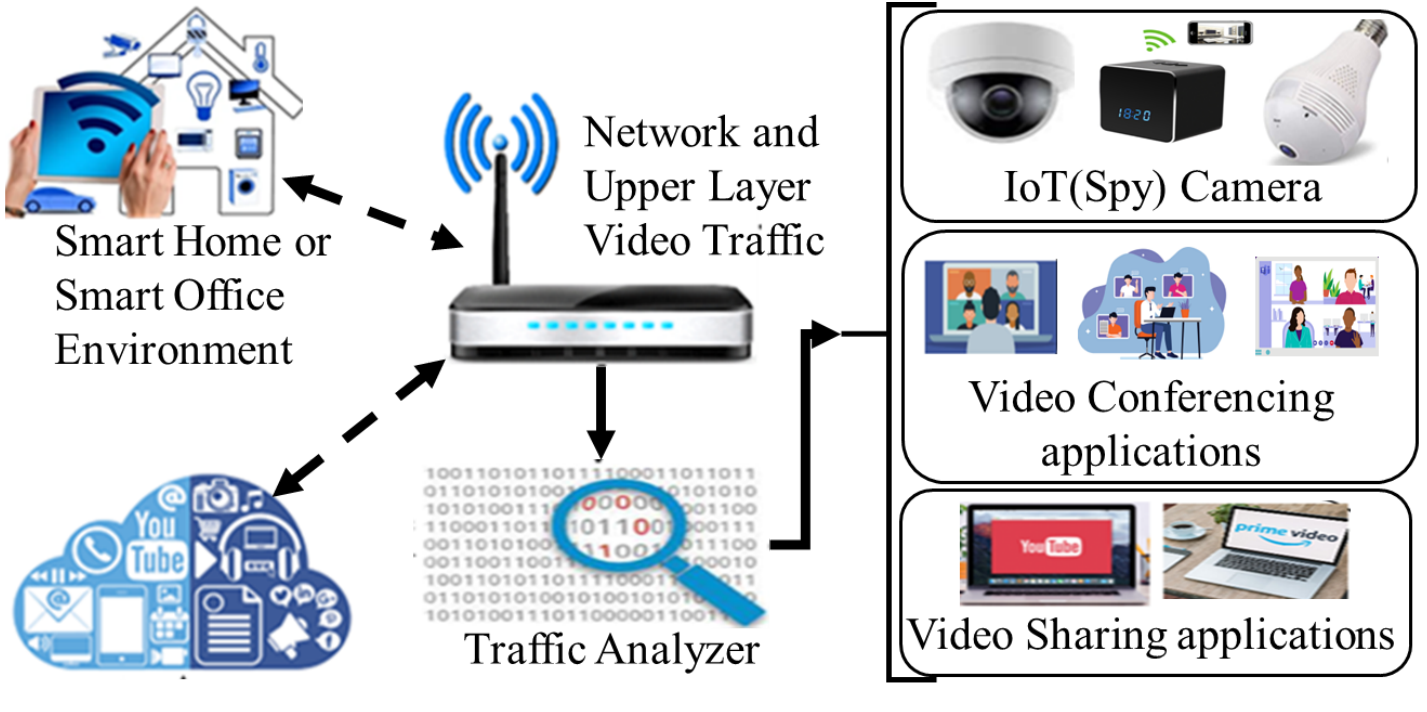}
  \caption{A motivating example for IoT (spy)camera installation.}
  \label{fig:motivate}
\end{figure}

We motivate our study by using a smart home scenario shown in Figure \ref{fig:motivate}. The smart home has a number of traditional computing devices, like desktop or smartphone, running certain online audio/video conferencing applications like Meet and video sharing applications like YouTube. 
Certain IoT devices, like smart bulb and smart door bell, are also allowed in the smart home. 
Installation of a multi-functional IoT device, like smart bulb cum spy camera (Spy Bulb) \cite{lightbulbcumcamera}, can create significant security and privacy threat if not authorised apriori. 
In this scenario, we ask two broad questions: i) is it possible to classify the traffic of conferencing applications from that of IoT camera? and ii) can the classification be agnostic of both the IP addresses and the port numbers especially when routers use NAT or applications use common cloud service providers \cite{Mazhar2020arXiv}? 

In this paper, we aim to build a system called \emph{iCamInspector} that can extract flow-based features from passively collected network trace using port mirroring on a switch for instance and classify traffic containing video data from IoT (spy) cameras in an environment where traditional online audio/video conferencing and sharing application can be executed. 
iCamInspector does not attempt to decrypt any network packet nor use IP address or port number as feature. 
We consider about 36.83GB of network traffic (basically pcap files) containing video data, divided into three broad groups based on the type of applications generating the traffic: 
(i) video traffic of four conferencing applications (Skype, Meet, Zoom and Teams), (ii) video traffic of two video sharing applications (YouTube and Prime), and 
(iii) video traffic of six IoT cameras.
The complete data set is collected in our own IoT laboratory.  
We have finally deployed iCamInspector in an unseen network scenario containing a set of IoT camera (not necessarily used before) and a set of conferencing applications that are used by some volunteers.

Our primary contribution in this paper lies in designing, developing and deploying our proposed system \emph{iCamInspector}. 
We summarise our contributions as follows.
\begin{itemize}
    \item We have developed and deployed \emph{iCamInspector} that can classify IoT camera traffic, in other words detect IoT camera, from that of conferencing and sharing applications, using only flow-based features. 
    
    \item iCamInspector, using simple decision tree classifier, achieves an average accuracy of more than 98.5\% with a miss classification rate of about 1.5\% in a 10-fold cross-validation, after pruning of 17 features out of a total of 77 features extracted by a well known feature extraction tool called CICFlowmeter \cite{cicflowmeter}.  
    
    \item We have achieved a reasonable performance of iCamInspector in a real deployment where IoT camera is detected with an average prediction probability of more than 0.9. 
    
\end{itemize}

In rest of the paper, 
Section \ref{sec:relwork} describes the related works. 
Section \ref{sec:background} briefs about the protocols for real time applications. 
Section \ref{sec:sysem-architecture} provides the design and implementation of iCamInspector.
Section \ref{sec:dataset} summaries the datasets used in this paper.
Section \ref{sec:conf-applications} \ref{sec:video-sharing} and \ref{sec:detection-iot-camera} investigates the protocols observed in conferencing applications, sharing applications and IoT camera respectively.
Section \ref{sec:cam-detection} and \ref{sec:ml-classifier} describes the features engineering and the performance of iCamInspector respectively, followed by conclusion in  
Section \ref{sec:conclusion}.

\section{Related Work}
\label{sec:relwork}
The studies that investigate the feasibility of classifying or detecting network traffic containing video data can be categorised into two broad types: one that focuses on video traffic in traditional audio/video applications, like Skype and Zoom, and the other that focuses on the detection of video stream from IoT camera among a set of IoT devices. 

\subsection{Works on Classifying Conferencing Applications}
The motivation of the works in the first category is two fold, one being the network bandwidth requirement and the other being the issues related to privacy and security. 
Two popular audio/video conferencing applications in recent past have been Skype and GTalk. 
A number of studies have focused in the area of detecting the network traffic that carry Skype data at different stages of a conference call including the network location of the hosts, e.g., behind NAT or Firewall \cite{baset2004arXiv, ptavcek2012analysis, Adami2009Springer}. 
Skype make use of RTP protocol for which the application can dynamically assign the transport ports in order to hide the presence of Skype hosts \cite{SkypeRTPCodec, SkypeRTP, FastRTPDetection} and there are works that try to detect Skype traffic without using such transport ports and IP addresses of the hosts running the Skype application \cite{Freire2008IEEE, bonfiglio2007ACM}. 
Further, the detection of network traffic is also achieved even when the Skype packets are encrypted using SSL/TLSv1.x \cite{sha2016Springer}.  
A survey on the detection of several online  audio/video conferencing application like Skype, GTalk, MSN VoIP and Yahoo can be found in \cite{Hugo2014IEEE}. 
Further, a fingerprinting framework have been proposed to identify the devices, like iPhone and Google Phone, that are executing Skype application among other traffic types like SCP, ICMP and UDP \cite{Radhakrishnan2015IEEE}.  
Finally, the feasibility of using flow-based features for the classification of Skype and GTalk traffic has also been investigated \cite{Skype_GtalkSkype}. 

\subsection{Works on Classifying IoT Cameras}
Both the users and the developers community of IoT devices in general and IoT cameras in particular are less careful about the security and privacy features of the devices \cite{alharbi2018IET} attracting a large research community in this area over a wide spectrum. 
We categorise the works relevant to this into two broad classes: exploiting security vulnerability and detecting/classifying IoT camera, in particular.

Certain case studies on specific IoT camera (like VibeX Onvif YY HD IoT camera \cite{Abdalla2020ISDFS} and Belkin NetCam camera \cite{tekeoglu2015IEEEICCCN}), IoT Hub (like SmartThingsHub \cite{fernandes2016IEEE} that can integrate IoT cameras) reveal that the risk of security breach is higher in the IoT cameras compared to other IoT devices. 
The frameworks like IoT-Praetor \cite{wang2020iot} is proposed to detect malicious behavior of IoT devices in general, whereas the frameworks like DecoyPort \cite{Kim2007Springer} and Siphon \cite{guarnizo2017siphon} is proposed to derive attackers intent. 

The classification of IoT devices with an aim of providing a kind of access control has been popular where the use of machine learning has been predominant. 
Detecting hidden IoT camera from wireless traffic \cite{wang2020iot, cowan2020detecting} and from network traffic  \cite{detectingIoT_IEEE2020, SivanathanA2018IEEE, Gordon2021IEEE} have been proposed. 
Such works have been extended to industrial IoT systems as well \cite{NawrockiM2020IEEE} and the results show that such extension is not trivial.  
Classification of IoT devices is also proposed via the decryption of encrypted Wi-Fi traffic \cite{Zhang2018ACM} which exploits the vulnerability in the security settings of the IoT devices. 
Use of application layer information to classify IoT devices has also been explored with the help of natural language processing(NLP) on the data \cite{thomsen2020smartlampsmartcam}.  
Overall, we propose for the first time to identify effective flow-based network traffic features for passive classification of network traffic containing video data arising from IoT camera from that of audio/video conferencing applications.

\section{Background}
\label{sec:background}
This section describes some the protocols that can be seen in the applications requiring real time data, like audio/video conferencing and video streaming. 

\subsection{RTP and RTCP} 
\textbf{RTP} (Real-Time Transport Protocol) \cite{ieee-rtp} provides the services to applications requiring real time data and operate mainly on User Datagram Protocol (UDP) in TCP/IP stack. 

\begin{figure}[hbtp]
\centering
\includegraphics[width=\linewidth]{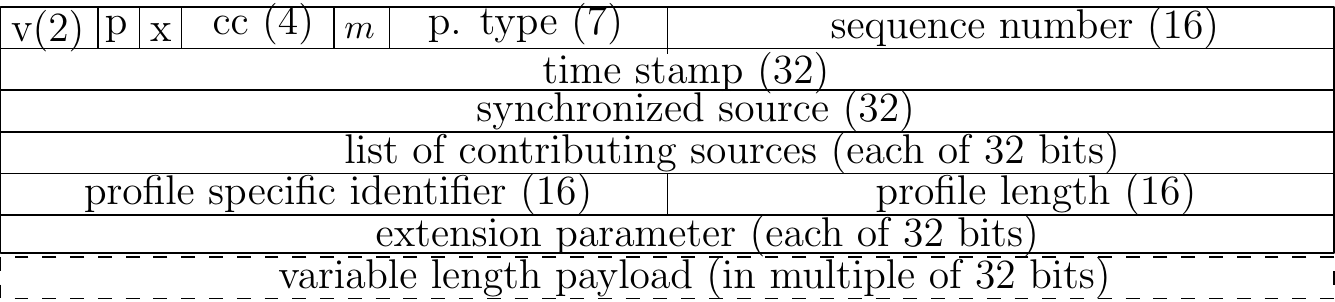}
\caption{RTP packet format, $x$ in ($x$) is number of bits; default is 1.}
\label{fig:rtp-packet-format}
\end{figure}

Unlike other transport layer protocols, a well known network packet analyzer, Wireshark \cite{wireshark}, does not automatically dissect the UDP payload as RTP packets. 
RTP packet format (Figure \ref{fig:rtp-packet-format}) provides a fixed header that specifies several mandatory information, like its version $V$ (currently $V=2$), the flags $p$ and $x$ indicate the presence of padding (often required for payload encryption) and the presence of header extension respectively.   
Contributing sources count $cc$ indicates the number of hosts present in the video stream and their identifiers are appended (in the \enquote{list of contributing sources}) after the fixed header portion of the packet.

The interpretation of the flag $m$ depends on the type of profile $p.type$, presently 128 profile types are supported.
The sequence number should be increased by one for every packet within a RTP stream and the time stamp indicates a sampling rate which is interpreted and utilized by upper layer. 
These two fields together offers a type of reliability on the RTP data to an upper layer protocol. 
All the packets in a RTP session have a unique source identifier called \emph{synchronized source} that adjusts timestamp in case multiple sources generates data. 
Note that RTP packet does not provide any provision of defining packet length, e.g., in bytes or words.

\textbf{RTCP} (Real Time Transport Control Protocol)
Lack of control mechanism over real time data stream in RTP demands control protocols, like RTCP \cite{ieee-rtp} and SDP (Session Discovery Protocol) \cite{Colin2006RFC}, that provides services like notifying to the active participants about the progress of the current stream. 
An application is free to choose the transport ports for both RTP and RTCP and hence not fixed. 

\begin{figure}[hbtp]
\centering
\includegraphics[width=\linewidth]{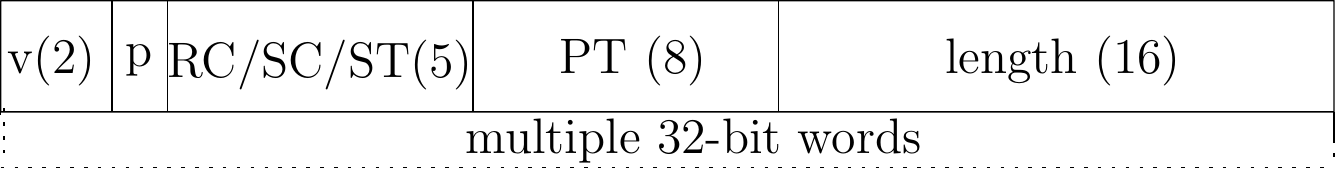}
\caption{RTCP packet format, $x$ in ($x$) is number of bits; default is 1.}
\label{fig:rtcp-packet-format}
\end{figure}

Similar to RTP, RTCP packet format (Figure \ref{fig:rtcp-packet-format}) has a fixed header followed by a variable part. 
First 32 bit in the packet contains version $V$ (currently $V=2$), padding $p$ flag, RTCP type (one of 200, 201, 202, 203 or 204) and a length field. 
The number of next 32-bit words depends on the RTCP type. 
Because the number of participants in a RTP session can be variable over time, the interval of RTCP packets can vary even within a same session (default interval is 5 seconds). 
We defer the discussion of  the protocol responsibilities for the later sections, and it is discussed if needed.

\subsection{QUIC -- Quick UDP Internet Connections}
\label{quicprotocol}
QUIC (RFC 9000 \cite{Jana2021RFC}) is relatively new protocol that is designed to overcome the performance challenges observed in TCP protocol. 
Recovery mechanism in QUIC can be used to implement reliable delivery and control congestion.
In general, UDP port number 443 indicates that the payload in the UDP packet is a QUIC packet. QUIC protocol defines two packet formats, long header (if first bit is 1) and short header (if first bit is 0) packets. 
Long or short header QUIC can be used by an application; we shall discuss further details of QUIC wherever required in this paper.

\subsection{UDPENCAP and ESP} 
\label{udpencapprotocol}
\textbf{UDPENCAP} (UDP Encapsulation of IPsec ESP Packets) \cite{Victor2005RFC} allows IPSec protocol to establish secure connection at Network Layer. 
Once such a connection is established, a host, rather an application within a host, can send encrypted packets to another host. 
Because underlying protocol is UDP, it is often suitable for the hosts to run multiple real time applications. 
The protocol has default support for NAT enabled hosts. 
The header of the protocol contains four fields (destination and source host, length and checksum), each of 16 bits followed by variable length ESP Header. 
Any host that make use of IPSec can be identified by the IP addresses used for enabling the IPSec protocol and hence we defer further discussion here in the section. 

\textbf{ESP} (IP Encapsulating Security Payload) \cite{Kent1998RFC} is one of the two protocols in IPSec that support security mechanism at Network Layer between two hosts, rather than applications running in the hosts. First two 32-bit words in the packet header contain security parameter index (SPI) and sequence number in the a security association.  Payload data field is variable in length for which padding is used to make it multiple of 32-bit words. 
The packet is then appended with authenticated application data. 
Because upper layer information is not available, we consider further discussion on the protocol later in this paper wherever required.

\section{System Architecture and Design}
\label{sec:sysem-architecture}
\begin{figure}[hbtp]
\centering
\includegraphics[width=\linewidth]{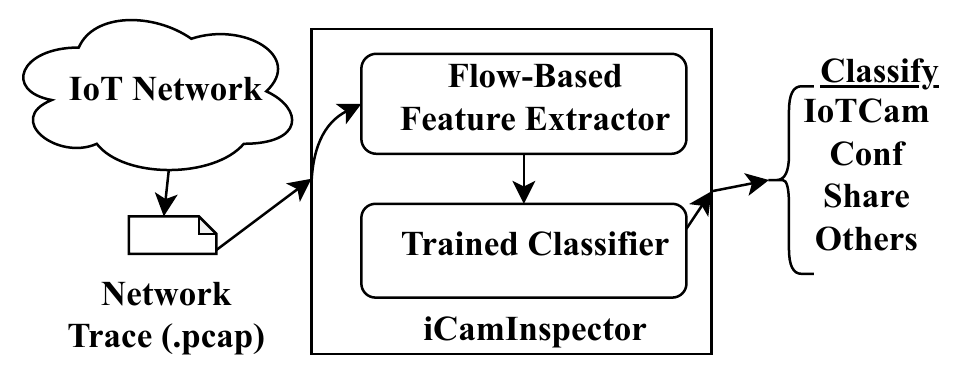}
\caption{Architecture of our project system called iCamInspector.}
\label{fig:architecture}
\end{figure}

Figure \ref{fig:architecture} shows the architecture  of our proposed system called \emph{iCamInspector}, which is a simple layered architecture. 
iCamInspector takes as input the network traces in the form of \emph{.pcap} files that can be collected by any standard tool like Wireshark \cite{wireshark} from a router or switch having port-mirroring option. 
Note that neither the router nor the switch restricts any end device to connect to it.
The flow-based feature extractor module takes \emph{.pcap} files as input and process it to compute certain properties. 
This module segregates a \emph{.pcap} file into a set of network flows. 
A flow is a 4-tuple of IP addresses and transport port numbers of two communicating hosts. 
This module then extracts a number of flow-based features (lets call a set of features a data point), from each of the flows. 
The Trained Classifier module then uses a set of data points corresponding to a set of network flows. 
The classifier is responsible for training and testing of ML models. 
This module uses four types of data points during training: IoTCam (indicating a flow  from one of the IoT camera), Conf (indicating that the flow is generated by one conferencing application), Share (indicating that the flow is generated by one video sharing application) or Others (indicating any other application). 
Thus, if an IoT camera is connected and become operational using such a router or a switch, then the camera can be detected by iCamInspector via classification of the network flows generated by the camera.  

\subsection{Threat Model}
We assume that an attacker can physically install an IoT camera with the help of stolen credential of a home Wi-Fi router, for instance. 
Alternately, a remote attacker can activate camera service in an IoT device that can offer multiple type of services, like smart plug-cum-camera or smart bulb-cum-camera.  
In general, we consider an IoT device as a spy device if it offers IoT camera service along with other service like bulb or plug. 
The attacker is not interested on the services like online audio/video conferencing that typically activated with the users knowledge. 
Once an IoT camera is installed or activated, it can stream the video to an attacker located remotely. 



\section{Experimental Setup and Data Set}
\label{sec:dataset}
\subsection{Implementation of iCamInspector}
We implement iCamInspector using freely available tools that are compatible with Python3. 
For the purpose of the paper, we have used a well known tool called CICFlowmeter \cite{cicflowmeter} to extract a total of 77 flow-based features from each flow, widely used SciKit Learn package for machine learning algorithms and Pandas for data preparation. 

\subsection{Network Setup for Data Collection}
how many devices connected to the same network? where are the applications running (e.g., laptop, Android, iOS)?

\subsection{Data Collection}
The data set in our experiments is a set of network packet capture (pcap) files that are collected when different Internet based application are executed in some clients or hosts. 
We segregate this network traffic data into three categories: Set I, Set II and Set III. 

In the Set I, four audio/video conferencing applications (Meet, Teams, Skype, Zoom) are used and a sum of about 10.06 GB of pcap files are collected from one of the participants in a conference call in each of the applications. 
The traffic is collected in two scenarios in each of the cases: with only audio and with only video is activated. 
In addition to the specific labels of Meet, Teams, Skype and Zoom to the traffic data of individual conferencing applications, we assign a collective label, "Conf", to this set of network traffic.  

In the Set II, two video sharing applications, YouTube and Prime Video, are used and a sum of about 11.81 GB pcap files is collected from the host downloading and playing the videos selected arbitrarily.
The traffic is collected when a video is downloaded and viewed live, i.e., we do not consider the videos that are already buffered or viewed before.  
In addition to the specific labels of YouTube and Prime to the traffic data of individual video sharing application, we assign a collective label, "Share", to the set of traces. 

In the Set III, six IoT cameras (Ezviz, D3D, V380 Spy Bulb, Netatmo, Canary and Alarm Spy Clock) are configured and accesses on a case-by-case basis by using different configuration settings, other than factory settings, and a sum of about 14.96GB of pcap files are collected from all these cameras. 
Note that, in this case, we collected network traffic from a router that connects the IoT cameras to Internet and this router was not necessarily used to connect our smartphone or Tablet to access the video stream from these  cameras. 
We consider only the traffic that a camera is associated with, and not the smart phone that views or controls the camera.
The IoT Cameras are configured by activating various services available in each of them, e.g., pan, tilt, 360 degree moving, switching audio on and off mode,  for collecting the traffic. 
In addition to the specific labels of Ezviz, D3D, Spy Bulb, Netatmo, Canary and Spy Clock to the traces of the individual IoT cameras, we assign a collective label, "IoTCam" to this set of network traffic.
A summary of the data sets is given in Table \ref{tab:camera-data-set}. 

\begin{table}
\caption{Summary of Dataset. Conf, Share, indicate the conferencing applications and video sharing applications resp.}
\begin{tabular}{|p{0.6cm}|p{1.0cm}|p{3.2cm}|p{0.7cm}|p{1.2cm}|}
\hline
Data Set & \multicolumn{2}{c|}{Applications/IoT Cameras} & Size (GB) & Packet Count\\
 \hline
 \hline
Set I & Conf & Meet, Teams, Skype, Zoom
 & 10.06 & 22,429,473 \\
\hline
Set II & Share & Prime Video, You Tube & 11.81 & 8,026,748\\
\hline
Set III & IoT Cameras & Ezviz, D3D, V380 Spy Bulb, Netatmo, Canary and Alarm Spy Clock  & 14.96 & 25,675,111 \\
\hline
\end{tabular}
\label{tab:camera-data-set}
\end{table}

\section{Inspecting Video-Conferencing Video}
\label{sec:conf-applications}

We expect audio/video conferencing applications to generate bi-directional video traffic and it need not be surprising. However, this analysis can help us to compare the characteristics of the video traffic of IoT cameras, because a same protocol, like RTP, QUIC, or UDT, can be used in both the types of applications. 
The clients in these applications are called peers as every participant in each of these applications can send video data.

Because the problem in this paper is to avoid the use of any IP addresses that are associated with an application, we focus on the payload of the transport layer packets, i.e., payloads of TCP or UDP packets, and we rely on the capabilities of well-known packet analyzer, Wireshark\cite{wireshark}, for decoding the packets at each layer. 

Unless certain heuristics are applied, Wireshark cannot decode UDP payload if the port number for a protocol in the UDP header is not standardised.   
For instance, RTP protocol is not assigned any particular port number, and therefore UDP payload carrying packets of such protocol cannot be decoded. 
And, we do not apply any kind of heuristic by default for decoding such payloads in this paper. 

\subsection{RTP in Skype}
Figure \ref{fig:UDPPorts_vShare}(a) shows a port-cloud to indicate a relative proportion of the port numbers seen in UDP packets in Skype traffic.  
Because Skype uses RTP for its audio/video data transfer \cite{bonfiglio2007ACM}, \cite{SkypeRTP}, we try to decode the UDP payloads having any of these port numbers. 
 
Based on RFC \cite{ieee-rtp}, we consider a valid RTP packet to have only version 2, otherwise it is not RTP/RTCP. 
RTP/RTCP packets can be multiplexed where both the protocols can use a same port number, as per RFC 5761 \cite{Colin2010RFC}, which is seen in Skype traffic. 

In Skype, RTP packet is separable from RTCP packet based on the values in bits from $3^{rd}$ to $8^{th}$ position in the UDP payload. 
The $3^{rd}$ bit is used for the indication of padding, which found to be 0 in both the protocols. 
The $4^{th}$ bit is set to 1 in all RTP packets, whereas it is 0 in all the RTCP packets. 
Bits from $5^{th}$ to $8^{th}$ (4 bits) indicates the contributing source identifier count in RTP, which is found to be either $(0000)_2$ or $(0001)_2$.
These four bits in RTCP have different types of interpretations, some of them are as follows. 
RTCP Feedback Message Type (FMT) Application layer feedback value denoted by $(1111)_2$; FMT Picture loss indication denoted by $(0001)_2$; Reception report count denoted by $(0000)_2$, $(0001)_2$ or $(0010)_2$; Generic negative acknowledgement (NACK) denoted by $(0001)_2$.
Our analysis shows that the interpretation of a same bit pattern, like $(0001)_2$, depends on the application using RTCP, and Wireshark apply certain heuristic to decode those. 

\begin{figure}
\centering
\begin{tabular}{ccc}
\includegraphics[width=0.3\linewidth]{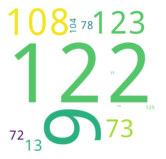} &
\includegraphics[width=0.3\linewidth]{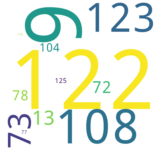}&
\includegraphics[width=0.3\linewidth]{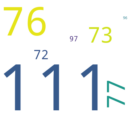}\\
(a) Skype & (b) Teams & (c) Meet \\
\end{tabular}
\caption{Word-cloud of RTP Payload Types in Conferencing Applications.}
\label{fig:AllRTPPayloadType}
\end{figure}

To distinguish RTP packets carrying audio and video data, we check second octet in the packet header.
It indicates \emph{marker} flag ($1^{st}$ bit), and the RTP payload type (next 7 bits in the octet). The Marker Flag values in video case are either 1 (True) or 0 (False), whereas in audio, its always 0 (False).
For instance, payload type $(0001001)_2$ indicates Dynamic RTP Type ITU-T G.722, which is Skype audio data;
whereas, $(1111010)_2$ or $(1111011)_2$ indicates Skype video data \cite{SkypeRTPCodec}. 
We have observed other payload types for both audio and video, but the proportion is different.
Figure \ref{fig:AllRTPPayloadType}(b) shows a codec-cloud seen both in audio and video. 
Video codec 122 is the highest among the observed codecs.

\subsection{RTP in Meet}
Decoding UDP payload as RTP (in a similar way as that in Skype) in Meet, we found that payload type field is $(1101111)_2$ and this denotes audio codec which is used by Google Hangouts for audio compression.  
The payload type is one of $(96)_{10}$, $(97)_{10}$, $(98)_{10}$, $(99)_{10}$ and $(100)_{10}$ when video data is carried in the packets. 

RFC 4856 \cite{Stephen2007RFC4856} and 3984 \cite{Thomas2005RFC} specify that these payload types can indicate any of audio and video codecs as these codecs are indicated by dynamic payload types that ranges from $(97)_{10}$ to $(100)_{10}$. 
An application can select any of them for audio or video. 

\begin{figure}
  \centering
\begin{tabular}{ccc}
\includegraphics[width=0.3\linewidth]{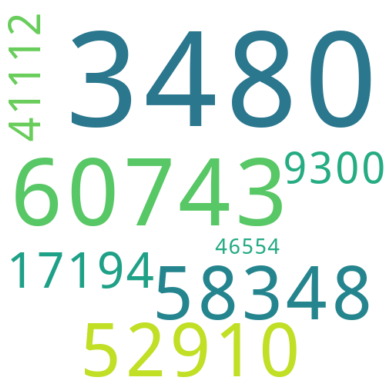}&
\includegraphics[width=0.3\linewidth]{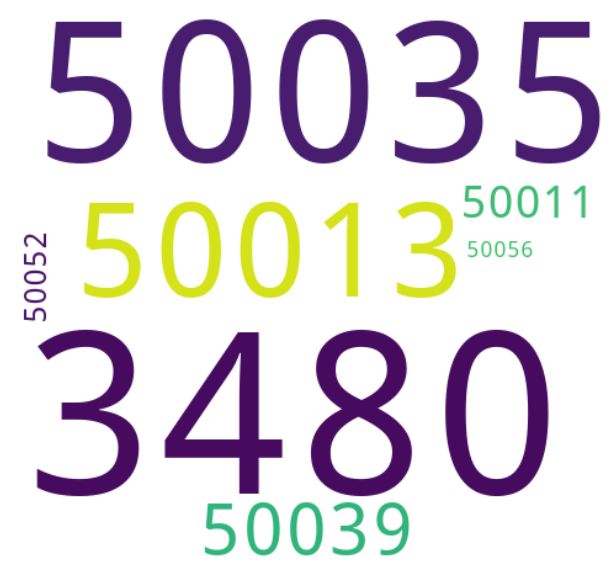}&
\includegraphics[width=0.3\linewidth]{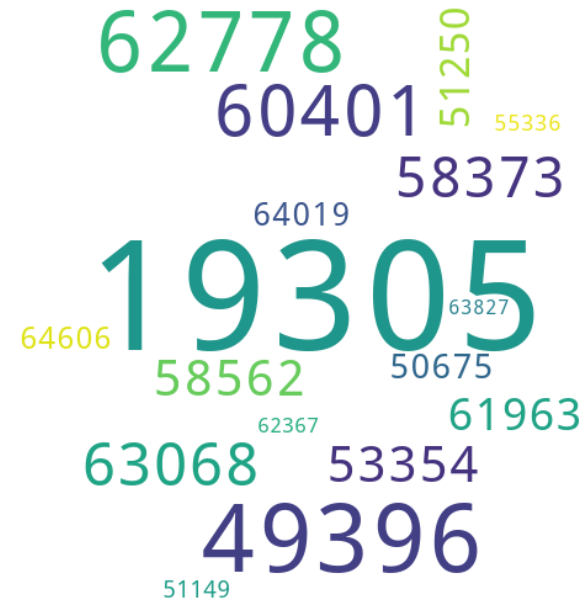} \\
(a) Skype & (b) Teams & (c) Meet \\

\end{tabular}
\caption{Word-cloud of ports in RTP in Conferencing Applications.}
\label{fig:UDPPorts_vShare}
\end{figure}

A large number of UDP port numbers is observed in Meet and, unlike others, the proportion of these port numbers (Figure \ref{fig:UDPPorts_vShare} (c)) indicates that port number 19305 is used with a much higher proportion compared to others in our data set.  
Further investigation during initial connection establishment shows that Meet involves a significant amount of DNS packets and each of the DNS packets is destined to Google servers.
Such DNS packets together with packets carrying audio/video data can distinguish Meet from others in the class of conferencing applications.

\begin{figure*}
  \centering
\begin{tabular}{cccccc}
\includegraphics[width=0.14\linewidth]{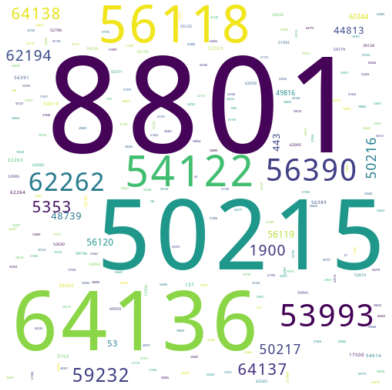} &
\includegraphics[width=0.14\linewidth]{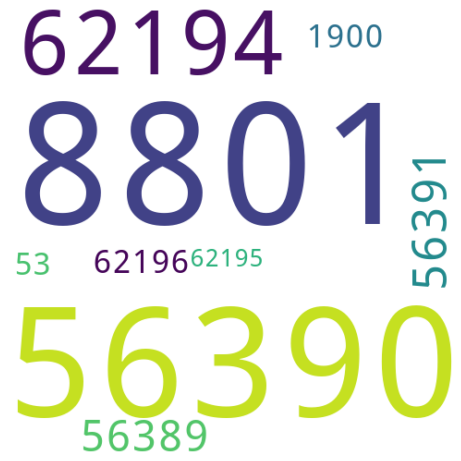} &
\includegraphics[width=0.14\linewidth]{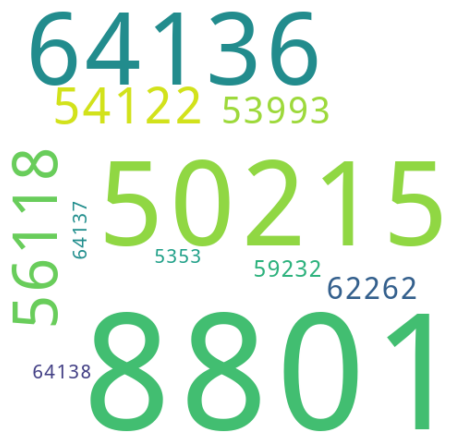}&
\includegraphics[width=0.14\linewidth]{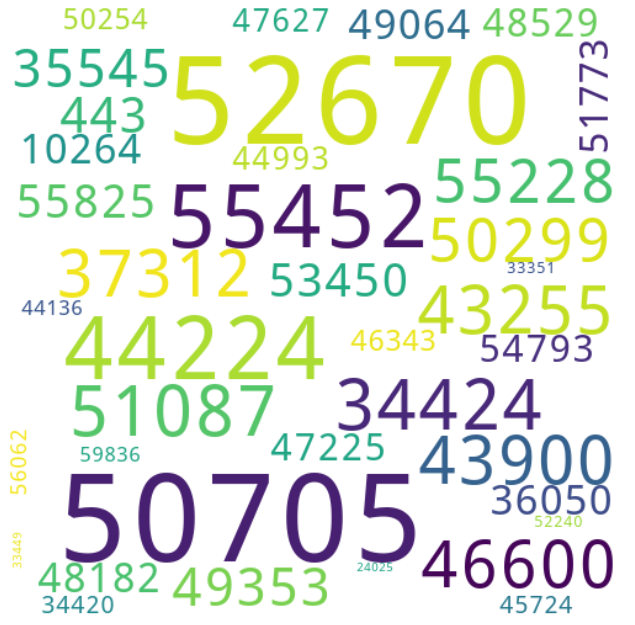} &
\includegraphics[width=0.14\linewidth]{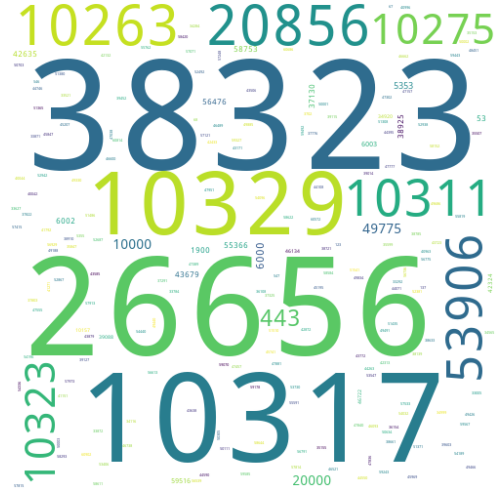} &
\includegraphics[width=0.14\linewidth]{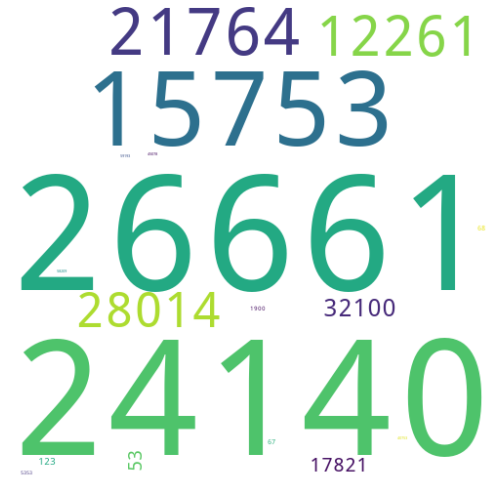} \\
(a) Zoom & (b) Zoom audio & (c) Zoom video & (d) D3D & (e) Ezviz & (f) Alarm Spy Clock\\

\end{tabular}
\caption{Word-cloud of UDP ports in Conferencing Applications and IoT Cameras.}
\label{fig:UDPPorts1}
\end{figure*}

\subsection{RTP in Teams}
In a similar way as than in Skype, we investigate Teams traffic. 
Teams uses multiple transport port numbers in RTP (Figure \ref{fig:UDPPorts_vShare} (b)). 
Comparing the RTP ports across these three applications, there is a little overlap in the sets of port numbers and hence a classification algorithm may not produce a good result that rely only on the port numbers.  

When Teams is used only with audio, RTP payload type contains $(104)_{10}$ or $(118)_{10}$ indicating audio compression codecs called  Silk and Comfort Noise \cite{MicrosoftCodec} respectively. 
Similarly, video traffic in Teams contains $(122)_{10}$ or $(123)_{10}$ that denotes video compression codecs known as H.264 (Advanced Video Coding) and H.264 FEC video compression codecs respectively. 
We correlated the payload types to the codecs based on Microsoft's web documentation \cite{MicrosoftCodec}.

We found that this conferencing application sends \emph{ping} packets to Microsoft servers during a call.
This together with RTP codes can confirm that a Teams call is in progress.

\subsection{UDP in Zoom}
We initiated our analysis of Zoom traffic where the video data is transfer through UDP protocol. Total UDP traffic is 93.78\% (989,854 packets out of 1,055,442), TCP traffic is 10.87\%  ( 114,713 packets out of 1,055,442) and TLS packets are 5.462\% (576,481 packets out of 1,055,442).
Because there is no documentation that we can find indicating Zoom uses RTP, we do not decode the UDP payloads as RTP. 
In fact, forceful decoding reveals RTP versions other than 2 and hence we do not consider that Zoom uses RTP. 

We find UDP port of 8801  to have a highest proportion (Figure \ref{fig:UDPPorts1}(a)); this is in-line with the findings in \cite{zoomudt2018Blog}.
Port 8801 is present in both audio and video traces of Zoom (Figure \ref{fig:UDPPorts1}(b) and (c) for audio and video resp.). 
Both audio and video traces contain TLSv1.2 packets only when a client connects to Akamai server.
In addition, the client often connects to Oracle corporations Server (193.123.197.43: 8801 or 52.202.62.252: 443) and sometimes to Amazon AWS server (3.208.72.69:443) during both audio and video calls. 
This is not surprising from the fact that Zoom has both Oracle Cloud and Amazon AWS as its partners for video data storage and infrastructure as service \cite{ZoomCloudAWSOracle}.

\section{Inspecting Video-Sharing Video}
\label{sec:video-sharing}
\subsection{YouTube}
Youtube uses TCP as its transport protocol. 
All the TCP packets that a YouTube client receives carry HTTPS payload (with server port as 443). 
A client uses several different TCP port numbers (Figure \ref{fig:TCPPorts_IPCamera}(b)).
About 90.5\% (accounts to 276218) packets belong to TCP, 21.29\% (accounts to 65000) of these packets are having length 5120 bytes or more. Overall, the TCP packet lengths vary in range 54--40936 bytes. 

Though we do not consider the domain names as a feature, YouTube client communicates with domains like \enquote{youtube-ui.l.google.com} (142.250.194.14:443, 142.250.77.238:443 or 142.250192.46:443) using TLSv1.3. 
In addition to it, we found domains like rr2.sn-ci5gup-h556.googlevideo.com (182.79.143.141:443) and rr3.sn-h557snzy.googlevideo.com (172.217.138.72:443). 
Worth noting that www.googlevideo.com is a reserve domain which provides free video hosting services so that the videos can be linked to www.youtube.com \cite{Vikas2016Blog, Hightech2022Blog}.

Importantly, we found QUIC protocol packets (with UDP port 443) at the beginning of playing a YouTube video or during advertisement associated with that video or during the loading the YouTube homepage itself. 
We have investigated in detail to see whether QUIC packets carry video data by blocking or unblocking the packets of this protocol at different stages of the video being played. 
our experiment reveals that QUIC packets only help in providing links to certain files, like video or advertisement. 
However, these packet do not carry video data. 
One of the primary reason for this conclusion is that the length of the packet in QUIC protocol is much smaller (67 -- 1292 bytes) and the fraction of these packets is also very less (about 0.6\% of the YouTube packets) compared to that with TCP port 443. 
Also, QUIC packets are not visible during the video being downloaded. 
This observation contradicts the observations in \cite{Youtube}, \cite{QuicYoutubeConnection}. 
And, we conclude that QUIC protocol is used to link the video or the advertisements within a video, and not to carry video data.

\subsection{Prime Video}
In Prime video traffic, about 99.9\%  (accounts to 4,009,009) of the packets belong to HTTPS (TCP port 443) (Figure  \ref{fig:TCPPorts_IPCamera}(a)).
We have seen a very few QUIC packets (0.002\% accounts to 1123) in this trace.
Like YouTube, QUIC protocol in Prime Video also uses UDP port number 443.  
Thus, we conclude that Prime servers transfer video data using HTTPS over TCP to its clients. 

In a similar analysis as YouTube,  we  have notices that about 78.85\% (accounts to 3,163,513) packets have length 1280 or higher bytes; 
maximum packet length is 1514 bytes. 
We have observed several domains, like 
www.aiv-cdn.net (13.35.231.97:443), www.akamai.net (175.101.127.217:443), www.llnwd.net (103.53.13.85:443), www.cloudapp.net (52.114.15.120:443), www.cloudapp.net (13.76.217.211:443), www.primevideo.com (13.35.212.192:443), www.media-amazon.com (13.249.229.167:443), www.cloudfrount.net (13.249.224.72:443) and www.llnwi.net (103.53.14.4:443); all of which communicate over HTTPS.

\section{Inspecting IoT Camera Video}
\label{sec:detection-iot-camera}

This section investigate the traffic of IoT camera with an aim of distinguishing it from the video traffic of conferencing and sharing applications for the first time in this paper. 

\subsection{Ezviz Camera}
Ezviz camera traffic has about 96\% (accounts to 5,727,256 packets) of UDP packets carrying application data, and rest belong to TCP, HTTP, DNS, TLS and QUIC to name a few. 
Several UDP ports (Figure \ref{fig:UDPPorts1} {(e)}) are seen while transferring application data with three ports (38323, 26656 and 10317) being prevalent. 
However, none of these three ports are assigned to any specific UDP-based application. 

Importantly, this camera directly interacts with both the mobile application and the cloud server (e.g., 47.241.58.2:6000 that belong to \emph{Alibaba (US) Technology}). 
We observed that about 5\% of packets are from camera to server, 8\% are from the server to the camera, 33\% are from the application to the camera and 54\%  are from camera to application. 
The packet lengths from camera to any of the application or the server vary in range [62, 1472] bytes; the length is 60 bytes from the server to the camera and the range is [54, 242] bytes from the application to the camera.

\begin{figure*}
  \centering
\begin{tabular}{cccccc}
\includegraphics[width=0.14\linewidth]{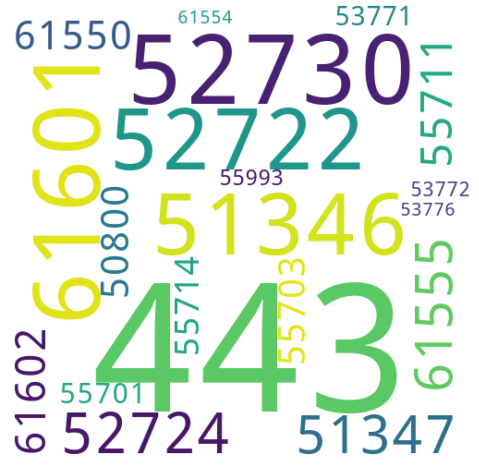}
&
\includegraphics[width=0.14\linewidth]{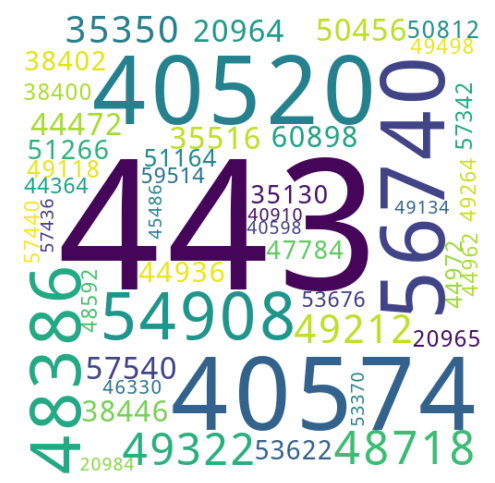} 
&
\includegraphics[width=0.14\linewidth]{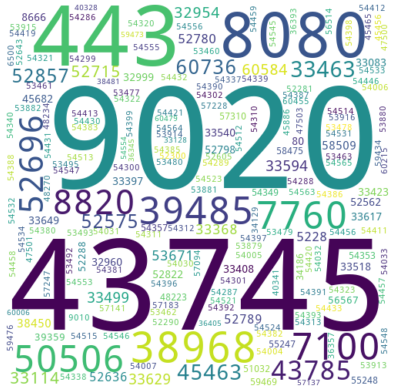}
&
\includegraphics[width=0.14\linewidth]{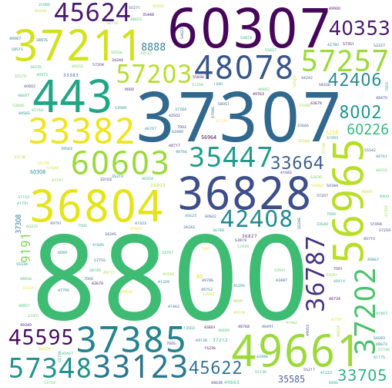}
&
\includegraphics[width=0.14\linewidth]{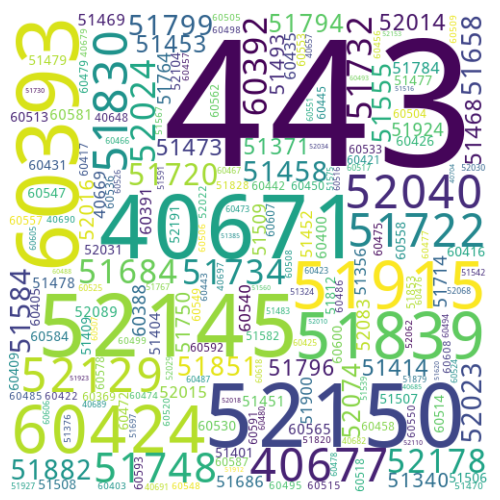} &

\includegraphics[width=0.14\linewidth]{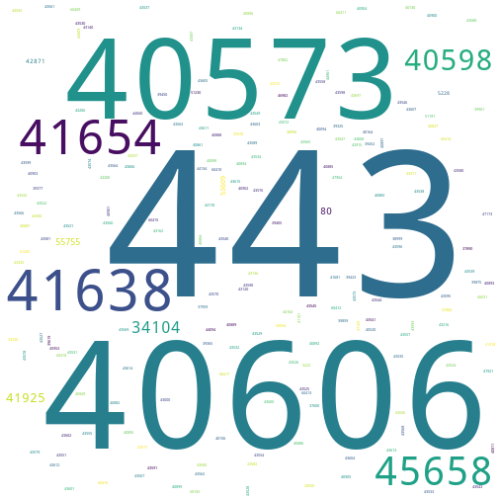}\\

(a) Prime & (b)  YouTube & (c) Ezviz & (d)  V380 Spy Bulb & (e) Netatmo & (f) Canary

 \label{fig:IPCam_TCPPorts}
\end{tabular}
\caption{Word-cloud of TCP ports in sharing applications (Prime and YouTube) and IoT Cameras (Ezviz, V380 Spy Bulb, Netatmo and Canary).}
    \label{fig:TCPPorts_IPCamera}
\end{figure*}

\subsection{D3D Camera}
D3D camera traffic has about 98\% of UDP packets, 0.7\% TCP packets and the rest belong to the protocols like ARP or DHCP. 
In UDP, about 1\% are DNS or NTP packets and the rest (about 98\%) carry application data.
A larger number of UDP ports (Figure \ref{fig:UDPPorts1}{(d)}) can be seen in this camera, and none of these is assigned to any specific application. 
Similar to Ezviz, this camera directly sends video data to both the mobile application and sometimes interacts with its cloud server (about 4\% packets). 
A server communicates using a fixed port to this camera and the port need not be same across the servers, e.g.,  47.94.37.184:32100 and 120.76.28.153:32100 belonging to \emph{Aliyun Computing Co. LTD. China}, and 47.89.48.227:32100 and 47.88.23.67:32100 belonging to 
\emph{Alibaba.com, China}. 
But, the camera and the app refit the port numbers frequently. 
The packet lengths from camera to any of the application or the server vary in range [46, 1074] bytes; the length is 60 bytes from the server to the camera (about 2\% packets) and the range is [46, 159] bytes from the application to the camera (about 35\% packets).

The camera communicates with some servers using HTTPS, e.g., 173.194.154.103:443, and some NTP servers (about 0.9\% packets) like 
\emph{www.ntp.org} (78.46.92.194:123), 
\emph{www.izatcloud.net} (52.42.72.58:123), 
\emph{www.ntp.org} (192.46.215.60:123) and  
\emph{www.nist.gov} (132.163.97.4:123). 

\subsection{Netatmo Camera}
Unlike the previous two cameras,  Netatmo camera does not directly deliver video traffic to its companion mobile application.  
Video traffic is sent to its servers (51.79.144.195:4500 and 51.79.156.58:4500) and then it is redirected from these servers to the mobile application.  
The traffic from the camera contains both TCP (about 9\%) and UDP (about 90\%) packets. 
RFC \cite{Victor2005RFC} and \cite{Kent1998RFC} indicate that UDP port 4500 can be used by IPSec protocol (particularly UDPENCAP and ESP). 
The camera mostly uses three UDP ports, i.e., 58275, 35301 and 36588. 

About 56\% of UDP packets containing application data are from the camera to the servers and the  lengths of these packets vary in range [43, 1422] bytes. 
The rest 34\% of UDP packets are from the servers to the camera and the length vary in range [90, 686] bytes.
About 3\% of the packets are TLSv1.2 and these packets are between the camera and the servers (like  
\emph{www.netatmo.net} -- 51.145.143.28:443) in both the directions.   
The length of TLSv1.2 packets varies in the range [54, 1514] bytes. 
Several other TCP ports are also observed (Figure \ref{fig:TCPPorts_IPCamera}(e)); prevalent are 40671 and 52150.

\subsection{Canary Camera}
Similar to Netatmo, Canary camera send video traffic to its servers and then the servers redirect the traffic to the companion mobile application. 
But, unlike Netatmo, this traffic contains about 99\% of TCP packets (of which about 57\% are TLSv1.2 packets) and the rest are UDP packets.
About 57\% of TCP packets are from the camera to the servers (of which about 99\% packets are TLSv1.2) and the rest are in reverse direction. The camera uses several TCP ports like 40606, 40573, 41638, 41654, etc. (Figure \ref{fig:TCPPorts_IPCamera}(f)).

More than ten cloud servers belonging to several domain communicate with the camera, a significant number of those servers communicate over HTTP, instead of HTTPS, e.g., 
\emph{www.conman.net} (82.165.8.211:80), 
\emph{www.canaryis.com} (34.230.191.96:80).  
In almost all the servers, the packet length vary in the range [54, 1514] bytes.

\subsection{V380 Spy Bulb}
Unlike other IoT cameras, V380 Spy Bulb offers two services: smart bulb and IoT camera. 
Because a single device offers two services, we have collected traffic in three scenarios in V380: i) only the bulb is operating, ii) only the camera is operating, and iii) both are operating.
In 'only bulb', several protocols are seen: 56.7\% ICMP, 21\% UDP application data, 16\% ARP, 5.3\% TCP, 1.4\% TLSv1.2 and the remaining are MAC authentication and DNS packets. Both TCP and UDP packets vary lengths in same roughly range of [50, 1514] bytes. 
In UDP, port 8877 is seen in about 90\% of the packets. 
TCP port of 880 remain constant in V380 for all TCP communications.
In 'only camera', 98\% of the packets belong to TCP and port 8800 is used in about 95\% of these packets 
(Figure \ref{fig:TCPPorts_IPCamera}{(d)}).
TLSv1.2 packets are about 0.2\% and we conclude that TLSv1.2 does not carry video data. 

In 'both' (i.e., bulb and camera), about 0.4\% are UDP packets (of which 0.3\% carry application data) and the rest are TCP packets. 
This essentially indicates that the packets carrying video data dominates. 
The port numbers and the packets lengths in both TCP and UDP are similar to that as we have observed in the first two settings. 

V380 interacts with a large number of domains, like \emph{Aliyun Computation Corporation Limited, China} (116.62.12.85:8888), 
\emph{www.av380.net, China} (120.24.33.45:8002), 
\emph{www.alibaba.com, Singapore}
(149.129.135.28:1340),  
\emph{www.nvcam.net, US} 
(47.74.159.35:8899), 
and \emph{Alibaba (US) Technology Co., Ltd., India} (149.129.135.251:443).
Only about 0.01\% packets (all of these are TCP) are from the camera to the server and the lengths of these packets vary in [66, 1514] bytes, whereas about 66\% packets (all of these are TCP) are from the camera to the application and the packet length vary in the same range. 

\subsection{Alarm Spy Clock}
Like V380, the Spy Clock sends video traffic directly to its companion mobile application. 
Unlike v380, this clock cannot be configured separately to avail only camera service or only clock service; both the services are clubbed together. 
The traffic contains about 99\% of UDP packets and the rest are TCP packets (out of which about 14\% is TLSv1.2 packets). Overall, about 71\% of the UDP packets are from the clock to the application and about 28\% are in reverse direction.  Several UDP ports can be seen in the Spy Clock, e.g., 26661, 24140, 12261, 21764 etc (Figure \ref{fig:UDPPorts1}(f)). 
UDP packet lengths from the clock to the mobile application vary in range [46, 1074], whereas the range is [46, 842] bytes in the reverse direction. 

About 0.2\% of packets (all are UDP) are from the clock to several servers, like  \emph{Aliyun Computing Co., LTD} (47.112.162.49:32100), \emph{dongjing2.xyx.wiki} (18.182.200.45:32100) and \emph{meidong2.xyx.wiki} (3.221.182.58:32100)
Also, the clock uses TCP to interact with some servers, like 
\emph{www.aliyuncs.com}
(47.75.19.167:80) and the packet length vary in range [54, 1514] bytes.

\subsection{Summary on IoT Camera Traffic Inspection}
We summarise the analysis of network traffic of six IoT cameras to see if transport and upper layer protocol feature can be enough to distinguish them. 
Table \ref{tab:IoTDevice_Sum} summarises the findings of the usage of protocol in all IoT cameras.  

Comparing these protocols with that in conferencing and sharing applications, we conclude the following. 
\begin{itemize}
    \item Video data of IoT cameras can be sent over both TCP and UDP without having any fixed set of ports.
    \item Unlike conferencing applications, none of the six IoT cameras use  protocols like RTP, QUIC or UDT that are designed specifically for real time data transfer.
    \item Some IoT cameras use a same cloud service provider and hence it can be  difficult to detect these video traffic using traditional rule-based intrusion detection systems like Snort or Suricata or Firewall.  
\end{itemize}

These observations call for machine learning based solutions for the detection and/or identification IoT cameras in a smart environment in order to project user privacy.

\begin{table*}[]
\caption{Summary of video traffic in IoT Devices. A * indicates more port numbers or servers with IP addresses seen.}
\label{tab:IoTDevice_Sum}
\begin{tabular}{|l|l|l|ll|ll|ll|l|l|}
\hline
\multirow{2}{*}{\begin{tabular}[c]{@{}l@{}}IoT \\ Devices\end{tabular}} &
  \multirow{2}{*}{Services} &
  \multirow{2}{*}{Dir.} &
  \multicolumn{2}{l|}{Pkts \%} &
  \multicolumn{2}{l|}{Device Ports} &
  \multicolumn{2}{l|}{\begin{tabular}[c]{@{}l@{}}Pkts len\\  Range (Bytes)\end{tabular}} &
  \multirow{2}{*}{\begin{tabular}[c]{@{}l@{}}Servers\\ IPAdd:Port\end{tabular}} &
  \multirow{2}{*}{\begin{tabular}[c]{@{}l@{}}App \\ IPAdd:Pot\end{tabular}} \\ \cline{4-9}
 &
   &
   &
  \multicolumn{1}{l|}{UDP} &
  TCP &
  \multicolumn{1}{l|}{UDP} &
  TCP &
  \multicolumn{1}{l|}{UDP} &
  TCP &
   &
   \\ \hline \hline
\multirow{2}{*}{\begin{tabular}[c]{@{}l@{}}Ezviz \end{tabular}} &
  \multirow{2}{*}{Video} &
  In &
  \multicolumn{1}{l|}{41\%} &
  2\% &
  \multicolumn{1}{l|}{20856*} &
  32954* &
  \multicolumn{1}{l|}{54-242} &
  54-923 &
  \multirow{2}{*}{47.241.58.2:6000*} &
  \multirow{2}{*}{192.168.4.6:38323*} \\ \cline{3-9}
 &
   &
  Out &
  \multicolumn{1}{l|}{59\%} &
  2\% &
  \multicolumn{1}{l|}{26656*} &
  38968* &
  \multicolumn{1}{l|}{62-1472} &
  54-1454 &
   &
   \\ \hline
\multirow{2}{*}{\begin{tabular}[c]{@{}l@{}}D3D\end{tabular}} &
  \multirow{2}{*}{Video} &
  In &
  \multicolumn{1}{l|}{37\%} &
  0.1\% &
  \multicolumn{1}{l|}{32793*} &
  60089* &
  \multicolumn{1}{l|}{46-159} &
  54--159 &
  \multirow{2}{*}{\begin{tabular}[c]{@{}l@{}}47.94.37.184:32100\\ 120.76.28.154:32100\end{tabular}} &
  \multirow{2}{*}{192.168.4.17:49353*} \\ \cline{3-9}
 &
   &
  Out &
  \multicolumn{1}{l|}{62\%} &
  0.1\% &
  \multicolumn{1}{l|}{49353*} &
  36891* &
  \multicolumn{1}{l|}{46-1074} &
  54--178 &
   &
   \\ \hline
\multirow{4}{*}{\begin{tabular}[c]{@{}l@{}}V380 Spy \\ Bulb \end{tabular}} &
  \multirow{2}{*}{Bulb} &
  In &
  \multicolumn{1}{l|}{21\%} &
  0.4\% &
  \multicolumn{1}{l|}{8877} &
  47791* &
  \multicolumn{1}{l|}{50-1514} &
  50-1514 &
  \multirow{4}{*}{\begin{tabular}[c]{@{}l@{}}116.62.12.85*:8888 \end{tabular}} &
  \multirow{4}{*}{192.168.4.7:8800} \\ \cline{3-9}
 &
   &
  Out &
  \multicolumn{1}{l|}{7\%} &
  0.6\% &
  \multicolumn{1}{l|}{8800} &
  47791* &
  \multicolumn{1}{l|}{50-1514} &
  50-1514 &
   &
   \\ \cline{2-9}
 &
  \multirow{2}{*}{Video} &
  In &
  \multicolumn{1}{l|}{0.1\%} &
  43\% &
  \multicolumn{1}{l|}{55436*} &
  8800 &
  \multicolumn{1}{l|}{50-1514} &
  66-1514 &
   &
   \\ \cline{3-9}
 &
   &
  Out &
  \multicolumn{1}{l|}{0.1\%} &
  57\% &
  \multicolumn{1}{l|}{55436*} &
  8800 &
  \multicolumn{1}{l|}{50-1514} &
  66-1514 &
   &
   \\ \hline
\multirow{2}{*}{\begin{tabular}[c]{@{}l@{}}Netatmo  \end{tabular}} &
  \multirow{2}{*}{Video} &
  In &
  \multicolumn{1}{l|}{34\%} &
  4\% &
  \multicolumn{1}{l|}{\begin{tabular}[c]{@{}l@{}} 36588*\end{tabular}} &
  \begin{tabular}[c]{@{}l@{}}40671*\end{tabular} &
  \multicolumn{1}{l|}{90-686} &
  54-1514 &
  \multirow{2}{*}{\begin{tabular}[c]{@{}l@{}}51.154.143.28:443\\(TCP)\end{tabular}} &
  \multirow{2}{*}{192.168.4.15:52150*} \\ \cline{3-9}
 &
   &
  Out &
  \multicolumn{1}{l|}{56\%} &
  5\% &
  \multicolumn{1}{l|}{\begin{tabular}[c]{@{}l@{}} 3530*\end{tabular}} &
  \begin{tabular}[c]{@{}l@{}}60424*\end{tabular} &
  \multicolumn{1}{l|}{43-1422} &
  54-1514 &
   &
   \\ \hline
\multirow{2}{*}{\begin{tabular}[c]{@{}l@{}}Canary \end{tabular}} &
  \multirow{2}{*}{Video} &
  In &
  \multicolumn{1}{l|}{0.1\%} &
  28\% &
  \multicolumn{1}{l|}{\begin{tabular}[c]{@{}l@{}}12261*\end{tabular}} &
  \begin{tabular}[c]{@{}l@{}}40606*\end{tabular} &
  \multicolumn{1}{l|}{46-842} &
  54-353 &
  \multirow{2}{*}{\begin{tabular}[c]{@{}l@{}}82.162.8.211*:80\\(TCP)\end{tabular}} &
  \multirow{2}{*}{192.168.4.17:40606*} \\ \cline{3-9}
 &
   &
  Out &
  \multicolumn{1}{l|}{0.1\%} &
  72\% &
  \multicolumn{1}{l|}{\begin{tabular}[c]{@{}l@{}}12261*\end{tabular}} &
  \begin{tabular}[c]{@{}l@{}} 41654*\end{tabular} &
  \multicolumn{1}{l|}{\begin{tabular}[c]{@{}l@{}}46- 1074\end{tabular}} &
  54-384 &
   &
   \\ \hline
\multirow{2}{*}{\begin{tabular}[c]{@{}l@{}}Alarm Spy \\ Clock \end{tabular}} &
  \multirow{2}{*}{\begin{tabular}[c]{@{}l@{}}Clock \\  Video\end{tabular}} &
  In &
  \multicolumn{1}{l|}{28\%} &
  0.001\% &
  \multicolumn{1}{l|}{\begin{tabular}[c]{@{}l@{}}26661*\end{tabular}} &
  60431* &
  \multicolumn{1}{l|}{46-842} &
  54-353 &
  \multirow{2}{*}{\begin{tabular}[c]{@{}l@{}}47.112.162.49*:32100\end{tabular}} &
  \multirow{2}{*}{192.168.4.16: 26661*} \\ \cline{3-9}
 &
   &
  Out &
  \multicolumn{1}{l|}{71\%} &
  0.001\% &
  \multicolumn{1}{l|}{\begin{tabular}[c]{@{}l@{}}17821*\end{tabular}} &
  60431* &
  \multicolumn{1}{l|}{46-1074} &
  54-384 &
   &
   \\ \hline
\end{tabular}
\end{table*}

\section{iCamInspector: Feature Extraction}
\label{sec:cam-detection}
This section describes the process of flow-based feature generation and feature engineering in this paper. 
The aim is to create certain supervised machine learning model for detecting and classifying IoT camera traffic.

\subsection{Flow-based Feature Extraction}
We use a well known tool called \emph{CICFlowmeter} \cite{cicflowmeter} to extract 77 flow-based features. 
A flow in the tool is defined by a 6-tuple of [FlowID, SourceIP, DestinationIP, SourcePort, DesitnationPort and Protocol], where FlowID is a flow identifier uniquely identifying the flow itself, SourceIP and DestinationIP are IPv4 address of the source and the destination host in a packet, SourcePort and DestinationPort are the transport port numbers at the source and the destination host in a packet, and Protocol is the Transport protocol indicated 'protocol' field in IP header in the packet. 

Typically, a flow in TCP starts and ends with a SYN and FIN packets respectively, and in UDP, it is based on a time duration. 
CICFlowmeter limits a flow in terms of a time window (defaults to 600s) in both of them. 

A record in the csv report file of CICFlowmeter contains a total of 84 columns, where first column indicates FlowID, next 5 columns indicate the flow parameters, the next 77 columns indicates 77 features extracted from the flow and the last column indicates a label of the flow. 
CICFlowmeter computes 73 thousand, 70 thousand and 10 thousand records for our data set in IoT cameras, conferencing applications and sharing applications respectively, totaling to about 155000 data points. 

\subsection{Flow-based Feature Pre-processing and Analysis} 
\begin{figure*}
  \centering
\begin{tabular}{ccc}
\includegraphics [width=0.33\linewidth]{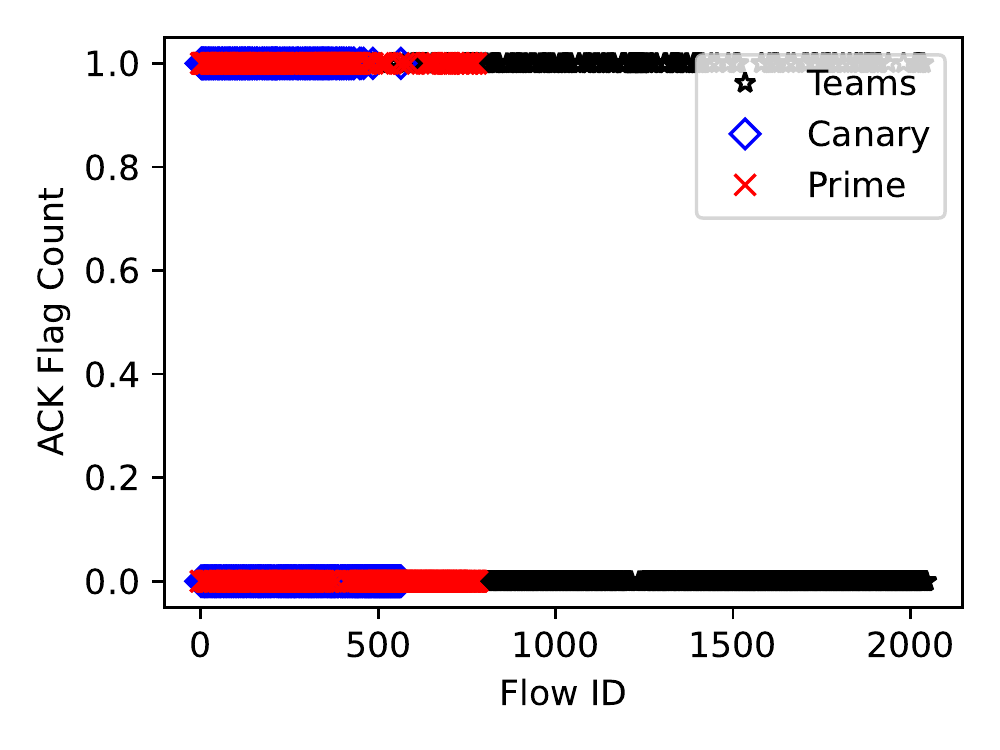} &

\includegraphics [width=0.33\linewidth] {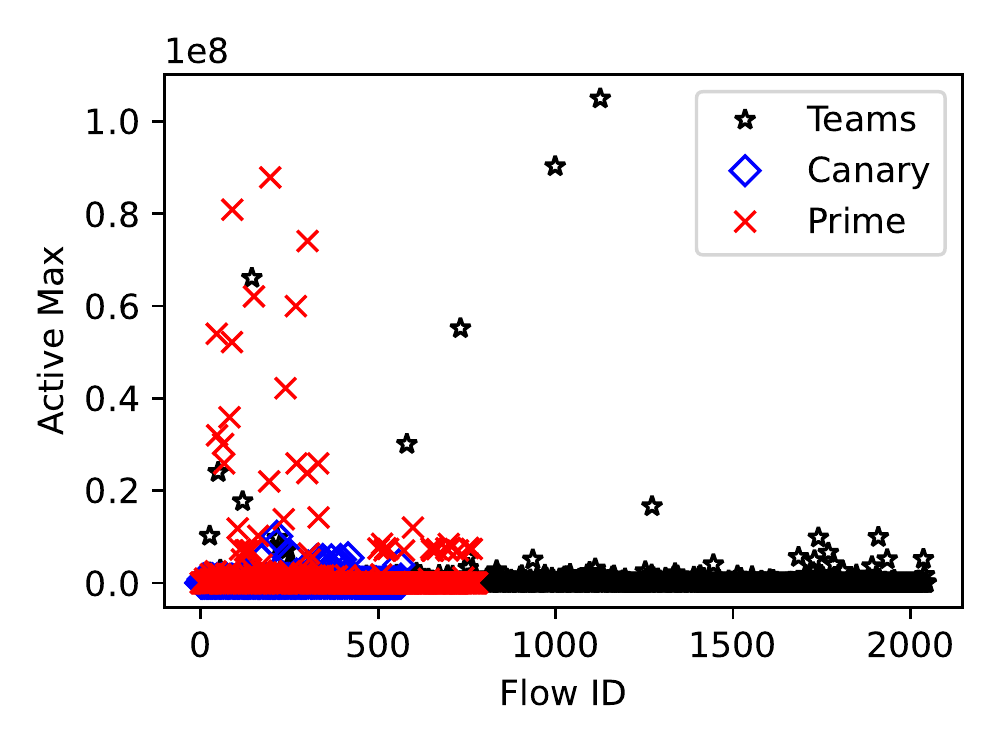} & 
\includegraphics [width=0.33\linewidth]{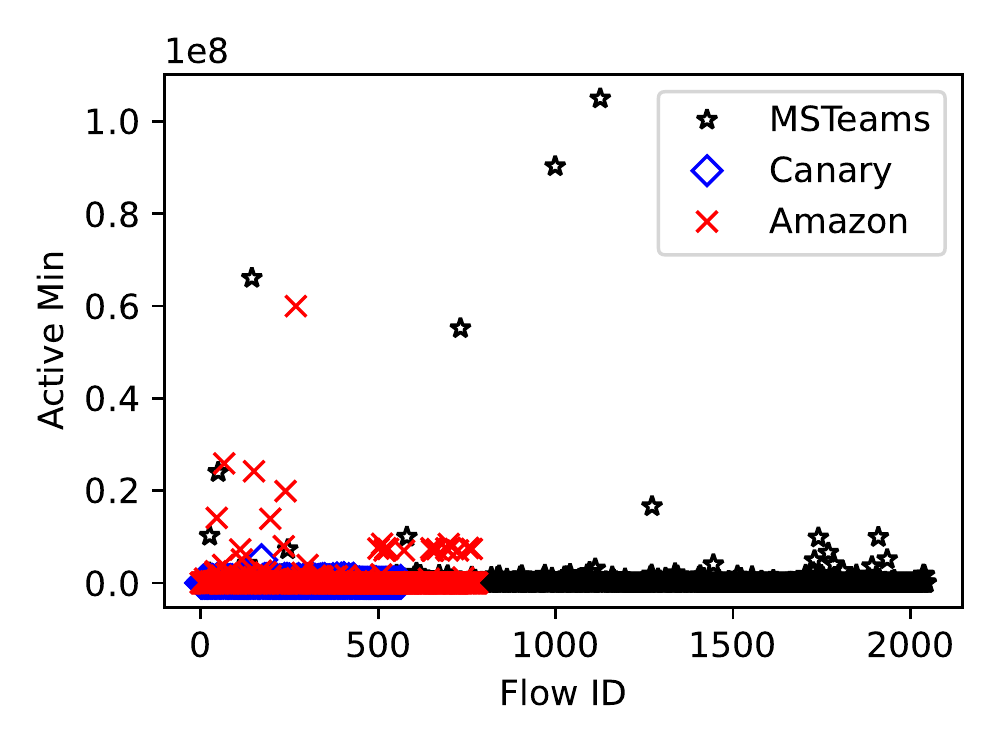}\\

(a) ACK Flag Count  & (b) Active Max  & (c) Active Min \\

\includegraphics [width=0.33\linewidth]{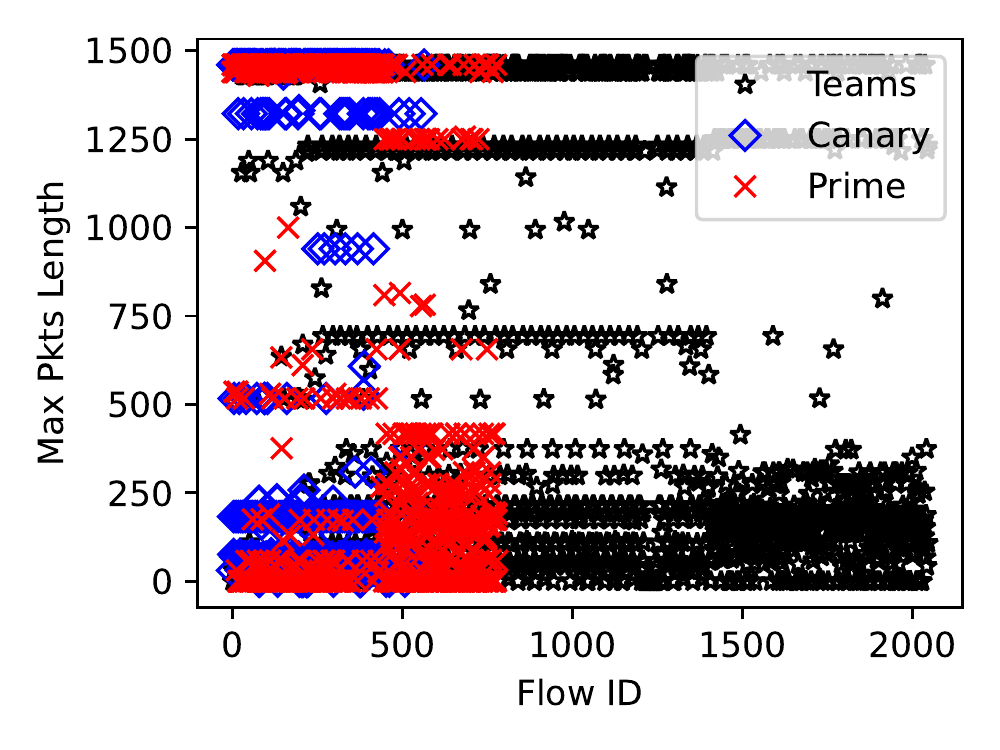} &
\includegraphics [width=0.33\linewidth] {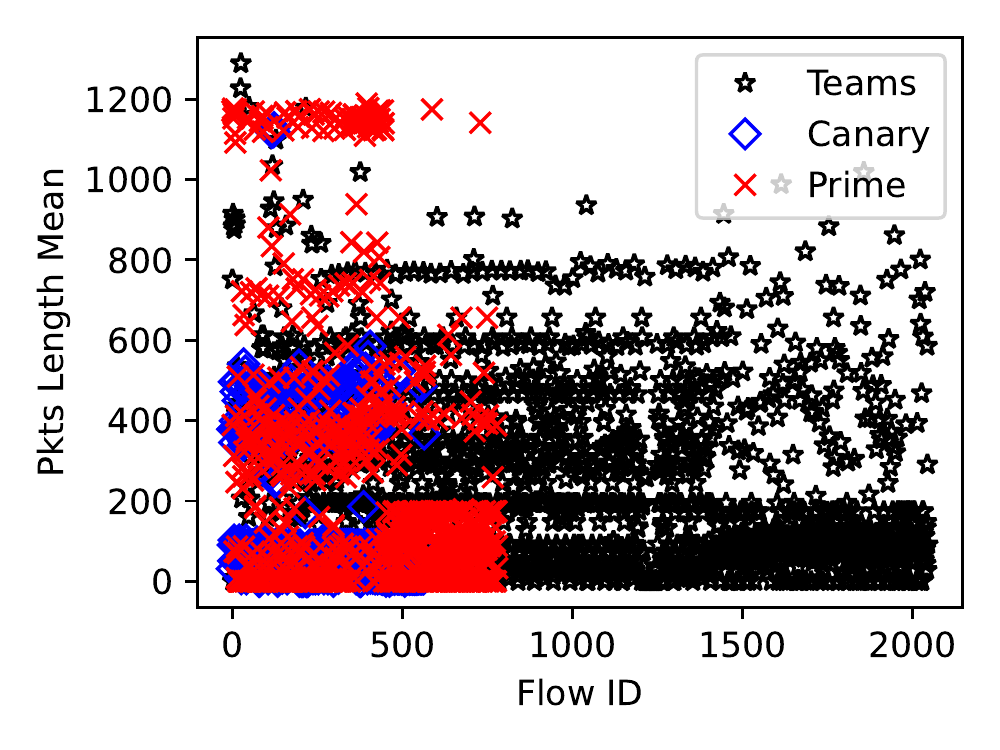} & 
\includegraphics [width=0.33\linewidth]{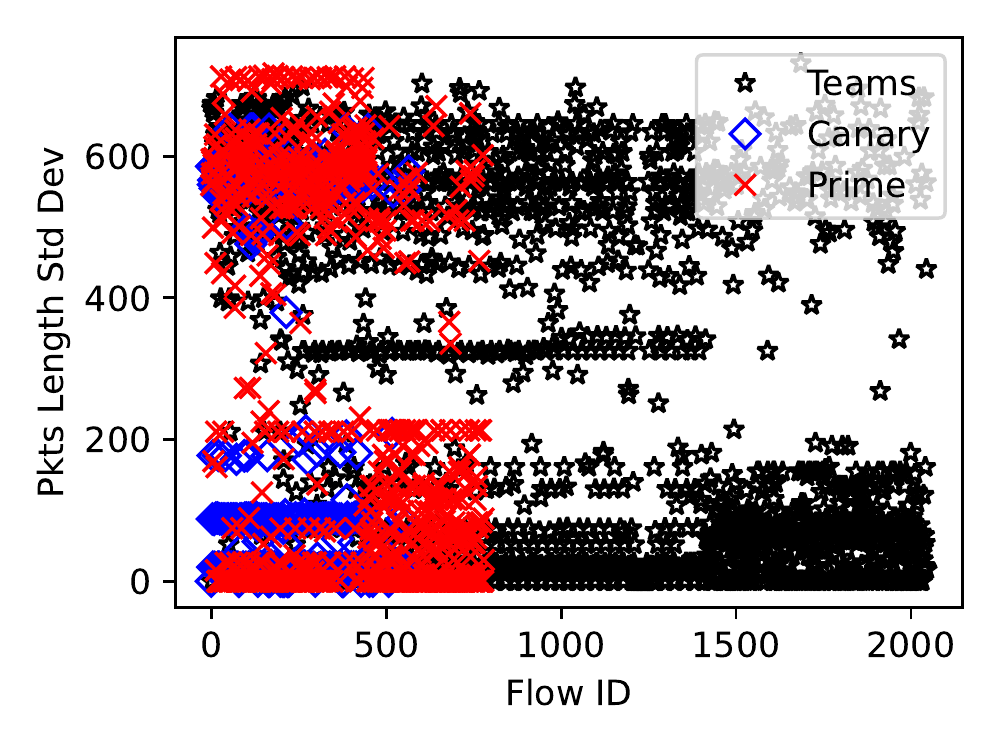}\\

(d) Max Pkt Len  & (e) Mean Pkt Len  & (f) Std Dev Pkt Len\\

\includegraphics [width=0.33\linewidth] {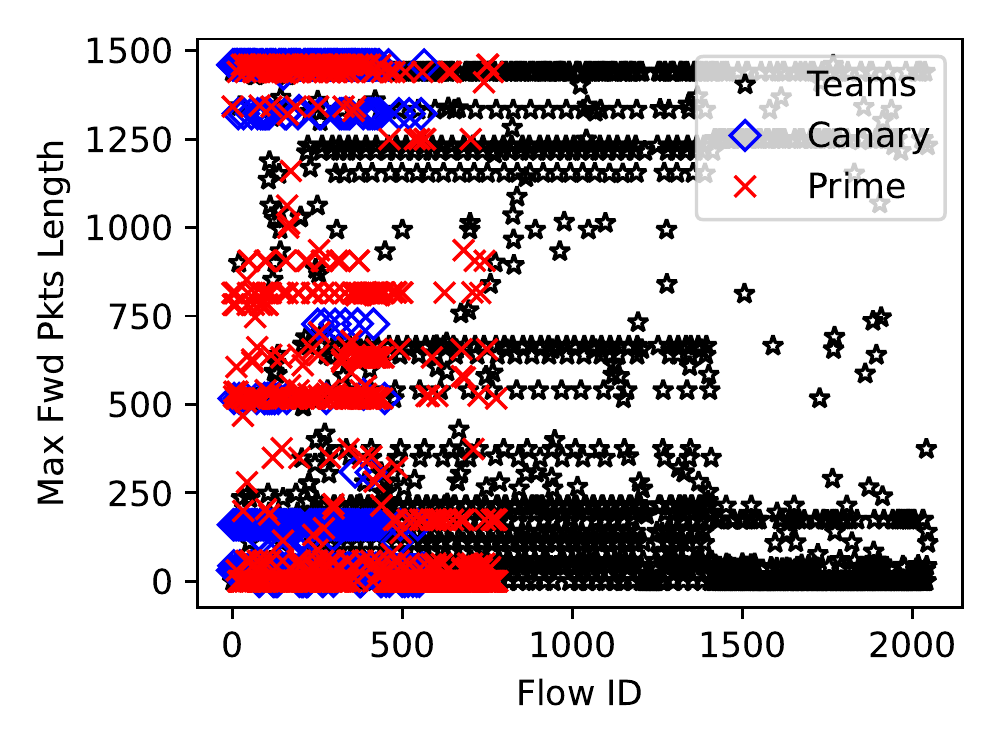}
&
\includegraphics [width=0.33\linewidth]{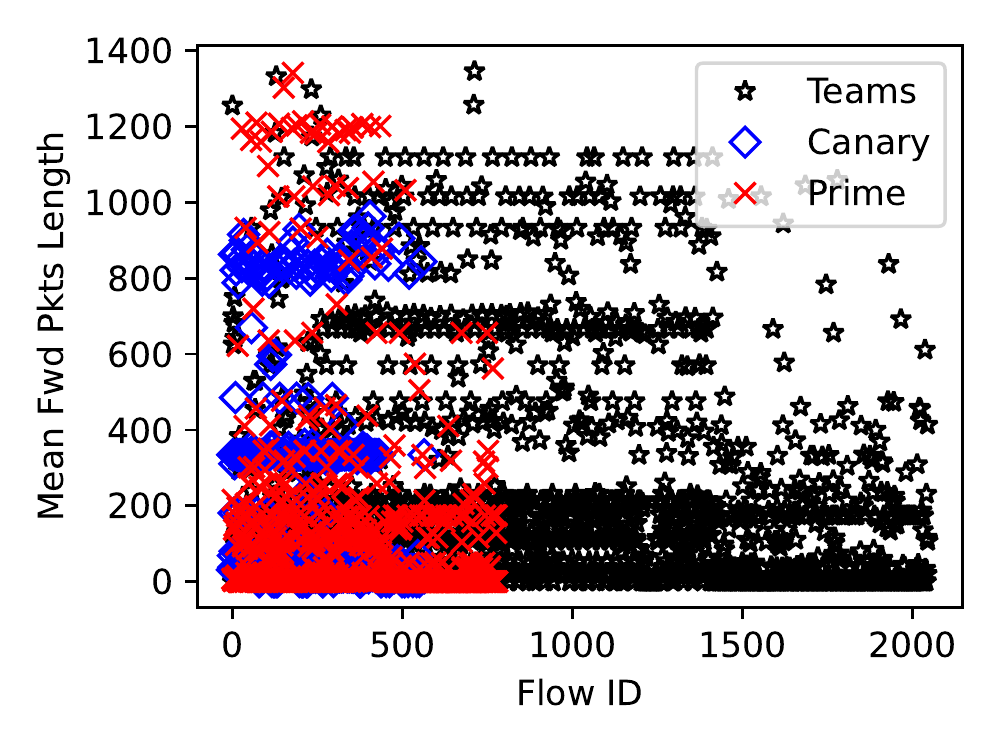} &
\includegraphics [width=0.33\linewidth] {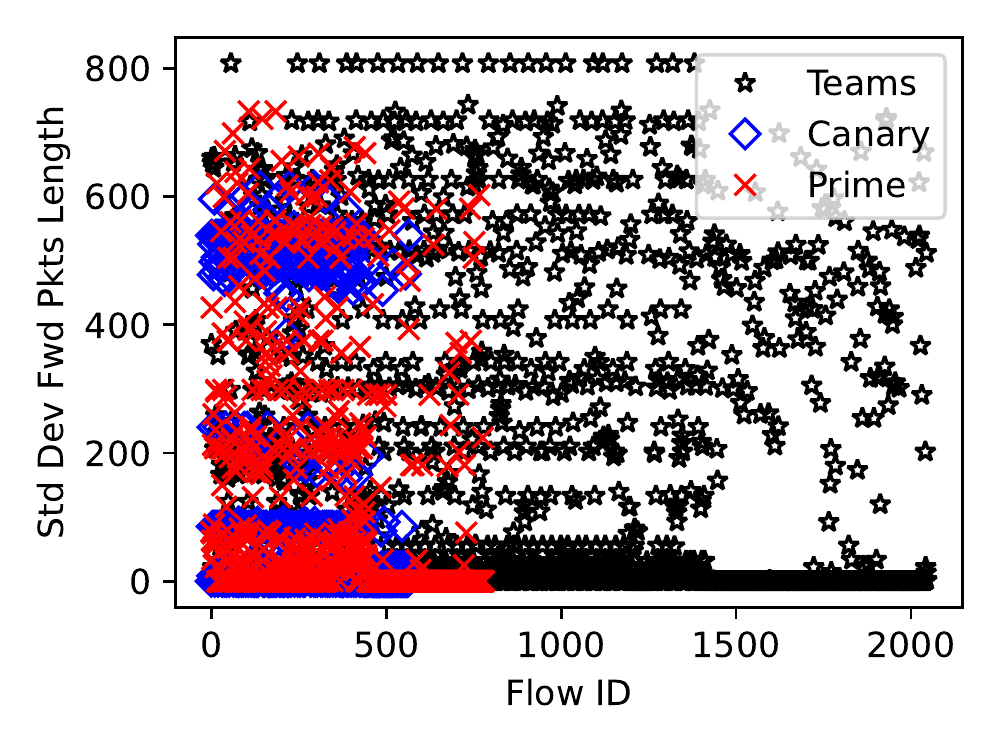} 
\\
(g) Fwd Pkt Len Max  & (h) Fwd Pkt Len Mean  & (i) Fwd Pkt Len Std Dev\\

\end{tabular}
    \caption{Flow Feature Plots in Teams, Canary camera and Prime applications.}
    \label{fig:flowfeaturesplots}
\end{figure*}

Based on an intuitive understanding of flow-based features, we aggregate the values of 12 features across the three types of applications and visualize them  for a comprehensive understanding. 
Figure \ref{fig:flowfeaturesplots} shows each of these features without any scaling in one of the subplots considering three arbitrarily chosen applications (Teams, Canary camera, and Prime video) from each of the categories. 
For a uniformity, we take a trace of about 15,50,000 packets in each of the three applications as input to CICFlowmeter. 

To start with, we check flow has at least one packet ACK flag On (Figure \ref{fig:flowfeaturesplots}(a)). 
It turns out that each of the applications has almost equal proportion of the number of flows having such packets or not. 
 
The maximum time duration that a flow is active before becoming idle (Figure \ref{fig:flowfeaturesplots}(b)) is relatively shorter in Canary camera compared to the others.

The threshold time gap between two successive packets that defines the boundary between active and idle state of a flow is internal to the CICFlowmeter and we do not attempt to configure it for the purpose of the work in this paper.   
The minimum time duration that a flow is active before becoming idle (Figure \ref{fig:flowfeaturesplots}(c)) also showing a similar trend with only exception that the range of this feature is slightly less in both Prime and Canary. 

The maximum packet length in each of the flows (Figure \ref{fig:flowfeaturesplots}(d)) varies in the range [34, 1460] in every application. 
While the values of this feature does not surprise much, it is a type of parity check that the packet length is constraint by the bandwidth of communication line. The mean of the packet lengths (Figure \ref{fig:flowfeaturesplots}(e)) in each flow turns out to be higher than 1200 bytes in Teams and Prime, whereas it is about 600 bytes in most of the cases in Canary. 
 
The standard deviation of the packet lengths (Figure \ref{fig:flowfeaturesplots}(f)) in Prime is relatively higher compared to that of the other two. The maximum of the packet lengths in the forward direction (Figure \ref{fig:flowfeaturesplots}(g)) can be viewed in three clusters, one in the range $[0,100)$, next in $[100,800)$ and the rest in $[800,1460)$.
The density of the data points in the cluster of $[0,100)$ is higher compared to other clusters in Teams. 
The spread of the data points in Canary is less compared to Teams and Prime.

Similarly, the mean of packet lengths in the forward direction (Figure \ref{fig:flowfeaturesplots}(h)) shows two clusters in Prime, one in the range $[0, 400)$ and the other in $[400, 1460)$ having centroids at 100 and 1000 respectively. 
In this feature, two centroids can viewed in Prime that are at 200 and 800, whereas three centroids can be viewed in Teams that are at 100, 600 and 1100. 
The standard deviation of the packet lengths in the forward direction (Figure \ref{fig:flowfeaturesplots}(i)) does not exhibit any such clusters in Teams and Prime. 

Looking further, we have observed that the maximum of packet lengths in the backward direction in Teams can be viewed in four clusters with the centroids at around 150, 650, 1200 and 1460; a similar structure of cluster can be seen in Prime and Canary.
However, the overlap among clusters is relatively less compared to that in the maximum packet length in forward direction.  
Such a distinction is difficult to visualize in the mean and the standard deviation of packet lengths in backward direction. (three plots on these three features are not shown due to page limitation).

\subsection{Summary of Feature Analysis}
With an aim of classifying network traffic originating  from IoT cameras, we have analyzed a set of flow-based features extracted. 
Note that no IP address is used as a feature in this paper.
One of the interesting and encouraging observation is that, given almost same number of network packets, the number of flows (Figure \ref{fig:flowfeaturesplots}(a)) created by Canary is much smaller than Prime and Teams. 
Combining the mean of packet lengths together with the number of flows, one can observe that the number of bytes transferred in each of the flows can allow us to distinguish the video traffic in IoT camera from that of other applications. 
Because not a single flow-based feature can clearly distinguish the video traffic, we sought for simple supervised machine learning models to our problem. 

\begin{table}[]
\caption{Summary of the number of records with label}
\label{tab:ml_dataset}
\begin{tabular}{|p{1.0in}|l|l|l|}
\hline
Applications & \#records & \#records/label & Label \\ \hline
\multirow{2}{*}{\begin{tabular}[c]{@{}l@{}}Prime\\  YouTube\end{tabular}} &
  \multirow{2}{*}{\begin{tabular}[c]{@{}l@{}}4314\\  6441\end{tabular}} &
  \multirow{2}{*}{10,755} &
  \multirow{2}{*}{Share} \\
             &           &                 &       \\ \hline
\begin{tabular}[c]{@{}l@{}}Meet \\ Teams\\ Skype\\ Zoom\end{tabular} &
  \begin{tabular}[c]{@{}l@{}}42493\\ 3953\\ 20947\\ 3464\end{tabular} &
  70,857 &
  Conf \\ \hline
\begin{tabular}[c]{@{}l@{}}Netatmo \\ Alarm Spy Clock \\ Canary\\ D3D \\ Ezviz \\ V380 Spy Bulb\end{tabular} &
  \begin{tabular}[c]{@{}l@{}}2775\\ 2278\\ 1937\\ 1819\\ 3064\\ 62119\end{tabular} &
  73,992 &
  IoTCam \\ \hline
\end{tabular}
\end{table}

\section{iCamInspector: Classifier Performance}
\label{sec:ml-classifier}

This section builds and evaluate the performance of the machine learning model upon the features engineered after getting computed by the CICFlowmeter. 
The aim is to see if a simple decision can be taken based on the inferences of flow-based features without relying on the domain names or the IP addresses. 
Thus, we consider simple decision tree based classification model for developing the classifier that can predict one among three classes of the video traffic. 
In other words, we investigate the performance of a decision tree based supervised machine learning model to detect if any IP camera is active in a smart environment where there is a possibility of running other traditional applications like video conferencing and sharing applications. 

\subsection{Building Supervised Classifier}
We begin with all 77 flow-based features that are extracted using CICFlowmeter. 
Table \ref{tab:ml_dataset} show a summary of the data that we use for building ML models. 
In brief, the number of records in the three classes 'share', 'conf' and 'IoTCam' are 10755, 70857 and 73992 respectively. 
Before we proceed to train ML model, the data set required certain cleaning and pre-processing. 
CICFlowmeter cannot generate a valid numeric values for each of the 77 features in each of the flows identified by the tool, possibly due to insufficient number of network packets belonging to the flow.
For instance, if the time duration is not at least 1s then computing the number of flow packets per second may produce NaN values. 
The number of columns having undefined values like 'NaN' or 'Inf' or '-Inf' is limited to one out of 77 columns in our complete data set.
And, the number of rows having such undefined values is less than 150 out of a total of about 153k data points.
Therefore, we choose to first replace all the 'Inf' or '-Inf' by 'NaN' and then replace 'NaN' by 0.

We consider decision tree classifier for its  elegance to comprehend about the features against the performance of the classifier. 
We focus on simple binary classification by considering three pairs of classes one at a time: ('IoTCam', 'Conf'), ('IoTCam', 'Share') and ('Conf', 'Share').

\begin{figure*}
  \centering
\begin{tabular}{ccccc}
\includegraphics[width=0.17\linewidth]{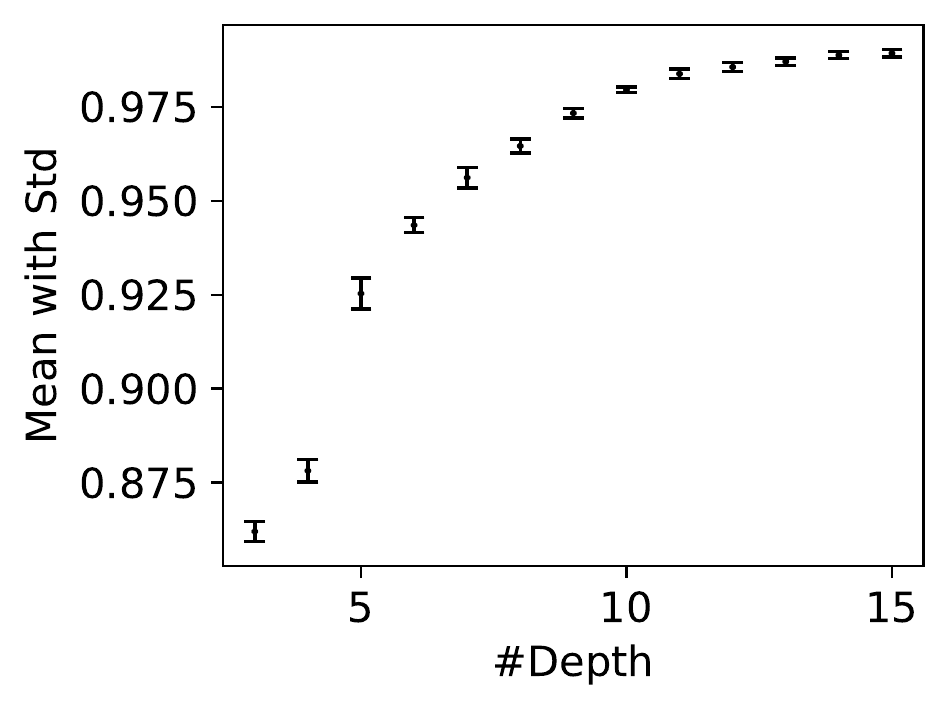}
&
\includegraphics[width=0.17\linewidth]{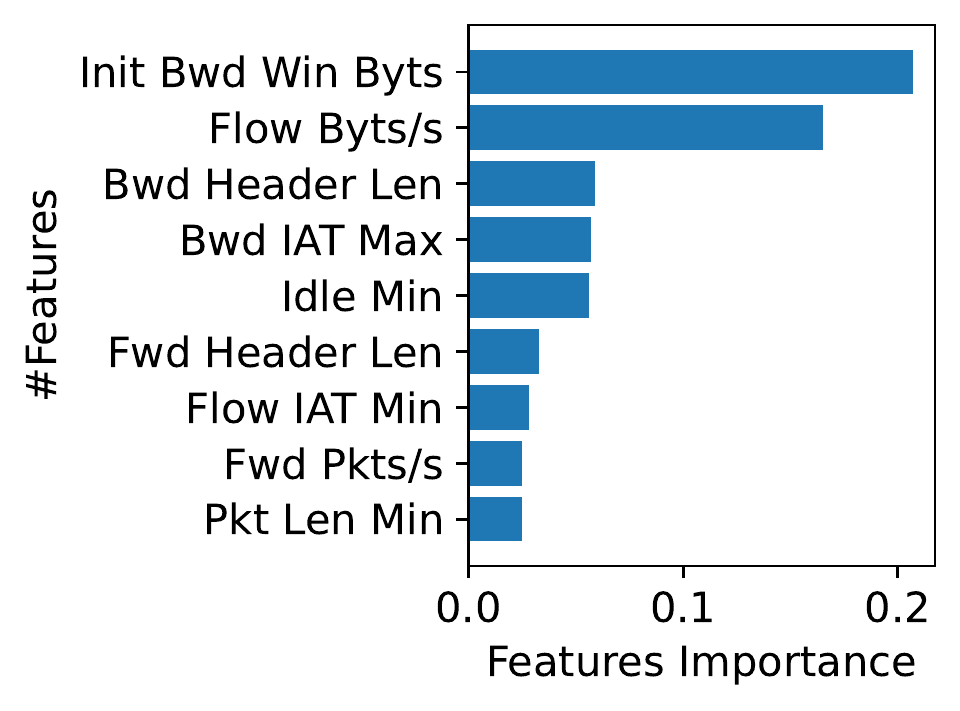} 
&
\includegraphics[width=0.17\linewidth]{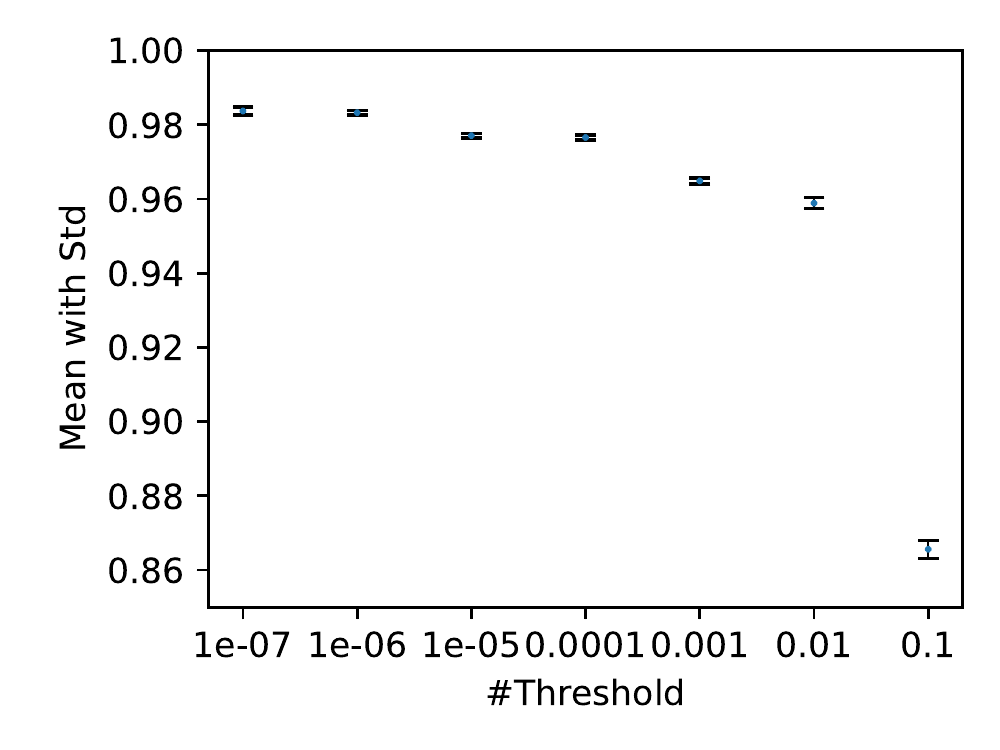}
&
\includegraphics[width=0.17\linewidth]{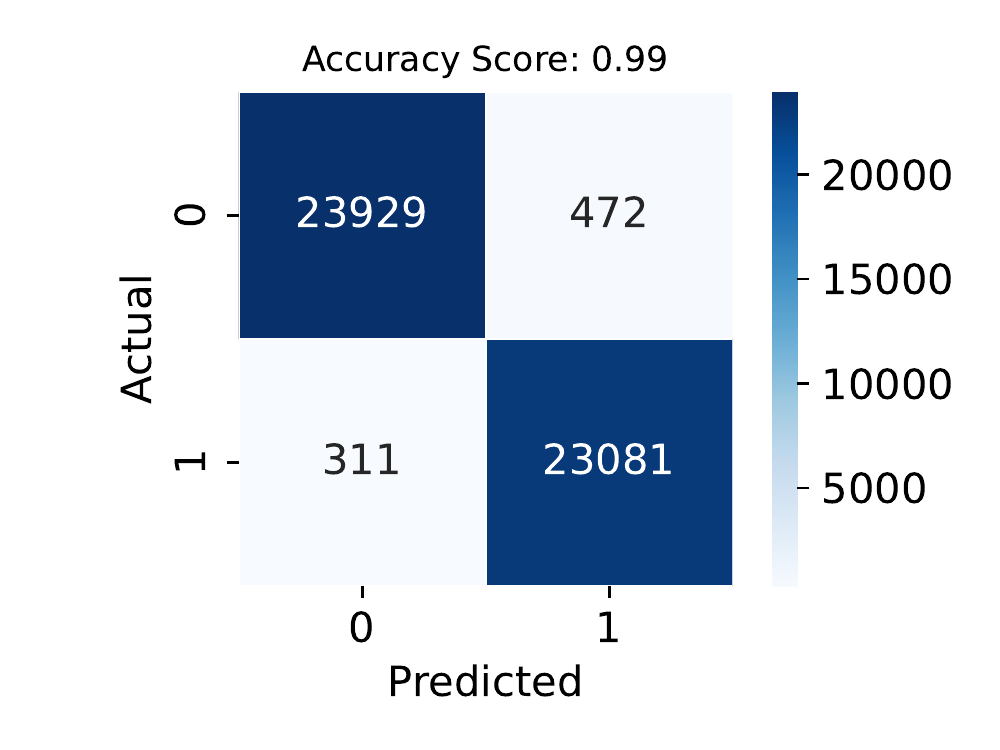}
&
\includegraphics[width=0.17\linewidth]{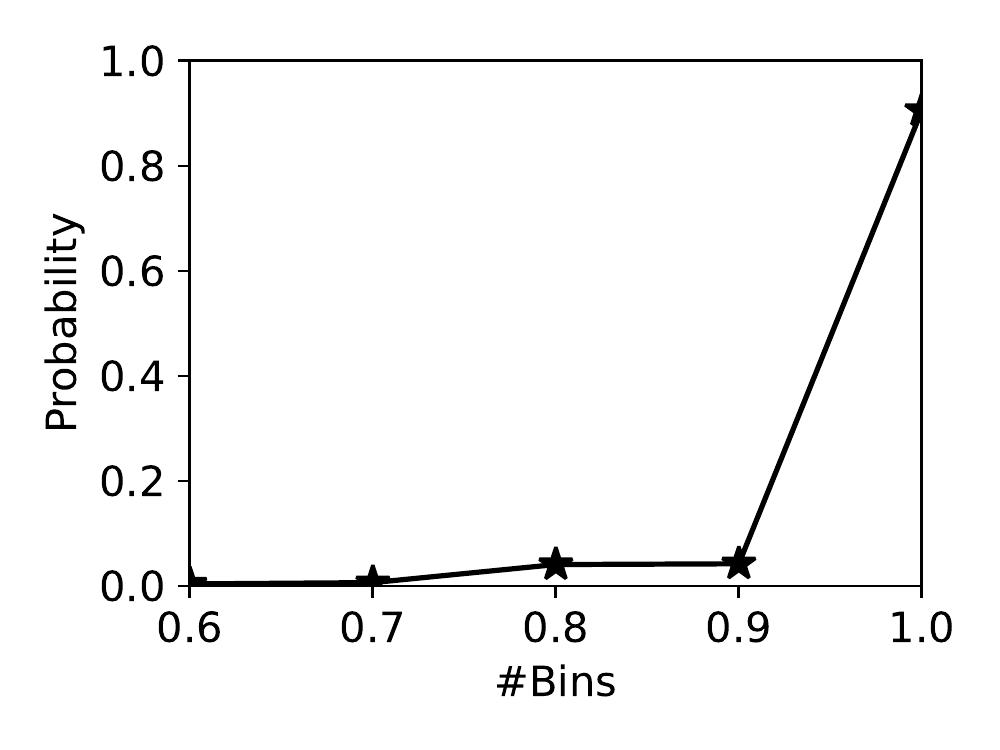} \\

(a) Mean of CV & (b) Feature Importance  & (c) Mean with threshold & (d) Conf. matrix & (e) Pred. Probability \\

 \label{fig:}
\end{tabular}
\caption{Results of iCamInspector classifier with deployment on real traffic (Cross Validation (CV), Confusion (Conf.) and Predicted (Pred.).}
    \label{fig:decision_tree_plot}
\end{figure*}

We perform k-fold ($k=10$ in this work) cross-validation for two reasons, one because data set is not quite balanced and the other because the number of exact applications in each class is limited (e.g., only two applications in the 'share' class). 
Figure \ref{fig:decision_tree_plot} (a) shows the mean of the cross-validation scores along with standard deviation with respect to the depth of the decision tree for the binary classification of 'IoTCam' and 'Conf'. 
For instance, the cross-validation produces an average of about 98.4\% accuracy with a standard deviation of about 0.001 with a maximum tree depth equal to 11.
Based on the cross-validation scores, we choose the tree depth as 11 for the next set of experiments. 

We look for features importance and found that 17 out of 77 features have negligible importance (less than 0.000001). 
Hence, we choose the rest of the 60 features to retrain a decision tree classifier and found that the model still achieves more than 98.0\% accuracy shown in Figure \ref{fig:decision_tree_plot} (b). 
So, for the rest of the experiments, we consider 55 features instead of 77 features with a tree depth of 11.

Figure \ref{fig:decision_tree_plot} (c) shows the mean of the cross-validation scores along with its standard deviation when feature important threshold increased from 0.0 to 0.1. 
The impact of increasing this threshold is in the reduction of the less important features.
We find that the threshold of 0.0001 creates 60 features and increasing the threshold further reduces the accuracy below 98.0\% and hence we fix to 60 features for rest of the experiments. 

Figure \ref{fig:decision_tree_plot} (d) shows the confusion matrix, where the miss classification of true 'IoTCam' predicted as 'Conf' is 1.32\% ($311/(23392)$) which is lower than the miss classification of 'Conf' as 'IoTCam' as 1.93\% ($472/(24401)$). 
Overall, we consider that 1.6\% miss classification rate is low. 
We surmise that the miss classified cases can be further associated with the specific DNS servers or the cloud servers or the NTP servers to identify their correct classes. 

Finally, we have used only the IoT cameras for classification, with a total of 14373 records (the data set is as shown in the last row of Table \ref{tab:ml_dataset} with only one exception that we have randomly chosen 2500 out of 62119 records in V380). 
We have used decision tree classifier with maximum depth as 11 with all 76 features. 
Our 10-fold cross-validation shows a mean accuracy of 80\% with standard deviation of less than 0.01\%.  
The false classification rate in this case is slightly higher, which is about 20\%. 
Analysis of feature importance shows that only 45 features have non-zero importance and these 45 features also achieves a similar accuracy. 
We conclude that the flow-based features extracted by CICFlowmeter may not be sufficient to classify the individual IoT cameras or a better classifier may be used to achieve a better performance.  
\subsection{Deployment of our IoT Camera Classifier}
Finally, we consider deploying our classifiers in real environment to further validate its performance.
In this case, we have collected about 2GB network traffic from a D-Link switch having port-mirroring option during a period of about 2 hours. 
We have activated a set of 3 arbitrarily selected IoT cameras, two of which are not considered in the other experiments in this paper, and requested two volunteers to access all these IoT cameras for some time in turns and do video conferencing with their friends. 
We have ensured that no other IoT devices or traditional computing devices are present in the network during the period of this data collection. 
We have used this traffic to extract flow-based features using CICFlowmeter and used 60 out 77 features that have been used for model development. 
Note that we do not have a ground truth to exactly validate the output of our trained classifier. 

Essentially, we have performed two-class classification in this case and the classes are 'IoTCam' and 'Conf'. 
iCamInspector detects 24401 (51.05\%) and 23392 (48.94\%) flows in IoTCam and Conf classes respectively out of a total of 47793 flows. 
We could roughly validate this result by inquiring the volunteers on how much time they spend on each of the applications. 
It turns out that the volunteers have spend their time in almost equal proportion in both video conferencing and interacting with the IoT cameras. 
They have frequently restarted their video conferencing with their friend and repeatedly activated and deactivated the associated mobile application to view the IoT camera feeds. 
Thus, because the number of 'Conf' flow is similar to that in 'IoTCam' and no other Desktop application and IoT devices were activated, we claim that classifier performs well in real deployment.
 
Further, we investigate the class probabilities for each of these flows that lead to the decision for the class prediction. 
Figure \ref{fig:decision_tree_plot} (e) shows more than 90\% of the flow have prediction probability in the range [0.9, 1.0], which we consider as a good performance of the classifier. 

\subsection{Limitation of the Approach}
While the performance of iCamInspector is commendable, we see a number of limitation of this approach. 
First, the data set should belongs to IoT camera is limited, particularly in the presence of a large number of commercially available IoT devices that offers multiple services.
For instance, a smart bulb that offers spy video streaming can significantly affect the performance if the video streaming data in not captured sufficiently. 
Second, after detecting the presence of IoT video stream, classification of the video stream and the non-video stream from a same IoT device may be challenging due to insufficient training data representing non-video streams. 
These limitations may lead to more exploration for the detection of IoT camera in a mixed environment and may sought for better classification models. 

\section{Conclusion}
\label{sec:conclusion}
In this paper, we have addressed a problem in cyber security domain of detecting IoT camera in a mixed environment, that have a higher risk of both vulnerability and privacy leakage compared to that of the other traditional online audio/video applications. 
We have designed, developed and deployed a system called \emph{iCamInspector} for the classification of IoT camera traffic from that of conferencing and sharing applications. 
Using six IoT camera from different vendors, two video sharing applications and four online audio/video conferencing applications, we have observed that the set of transport ports and the set of protocols above transport layer need not be fixed to any real transfer protocols like RTP, UDT or QUIC in the IoT cameras. 
Therefore, we resort to use a set of suitable flow-based network traffic features to classify the video traffic out of IoT camera from that in online conferencing and sharing applications. 
We show that iCamInspector that rely on a simple decision tree based classifier can achieve more than a mean accuracy of 98\% with standard deviation as low as 0.001\% in cross validation. 
Finally, the deployment of iCamInspector in an unknown network environment with a total of five video application including three IoT camera shows a commendable performance. 
We have observed that iCamInspector has certain limitation which may arise due to diversity of IoT camera and integration of more than one services in a single device, and hence may sought for further investigation.


\ifCLASSOPTIONcompsoc
  \section*{Acknowledgments}
\else
  
  \section*{Acknowledgment}
\fi

The authors would like to acknowledge the support the IoT devices that are part of a project funded by CRG-SERB-DST, Govt. of India, Project Code CRG/2020/005855.

\ifCLASSOPTIONcaptionsoff
  \newpage
\fi

\bibliographystyle{IEEEtran}

\bibliography{references}

\begin{IEEEbiography}[{\includegraphics[width=1in,height=1.1in,clip]{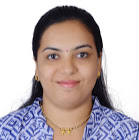}}]{Priyanka Rushikesh Chaudhary}
is working toward the PhD degree from the Department of Computer Science and Engineering, Birla Institute of Technology and Science Pilani, Hyderabad Campus, India, since 2019. Before that, she worked as an Assistant Professor at Dr. D. Y. Patil Institute of Management and Research, Pune. 
Her research interests are cyber security in Internet of Things.
\end{IEEEbiography} 
\vskip -2\baselineskip plus -1fil
\begin{IEEEbiography}[{\includegraphics[width=1in,height=1.25in,clip]{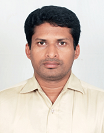}}]{Rajib Ranjan Maiti}
is currently an Assistant Professor in CSIS,  BITS Pilani, Hyderabad campusn India. He has done his PhD in CSE at IIT Kharagpur, India. His research interest lies in the area of Cyber Security in IoT and CPS. He has published his research works in journals like Transaction on Mobile Computing, Computer Networks and Cybersecurity, and conferences like Esorics, WiSec and AsiaCCS. He is currently executing two sponsored projects related to cyber security, one funded by SERB, DST, India and the other funded by Axiado, India.
\end{IEEEbiography}

\end{document}